\documentclass[showpacs,preprint,aps]{revtex4}
\usepackage{bm}
\usepackage{amsmath}
\usepackage{amssymb}
\usepackage{graphicx}
\usepackage{yfonts}\usepackage{mathrsfs}\include{mdwslides}
\include{fltfonts.def}
\usepackage{bm}\usepackage{epstopdf}\usepackage{float}\usepackage{mathrsfs}\usepackage{dcolumn}
\usepackage{bm}
\usepackage{mathrsfs}

\makeatother

\begin{document}

\title{Short Range Interactions in the Hydrogen Atom}

\author{A. D. Bermudez Manjarres }

\email{ad.bermudez168@uniandes.edu.co}


\author{D. Bedoya Fierro}

\email{da.bedoya528@uniandes.edu.co }


\author{N. G. Kelkar}

\email{nkelkar@uniandes.edu.co}


\author{M. Nowakowski}

\email{mnowakos@uniandes.edu.co}


\affiliation{Departamento de Fisica,\\
 Universidad de los Andes, Cra.1E No.18A-10, Bogota, Colombia }

\date{\today}
\begin{abstract}
In calculating the energy corrections to the hydrogen levels we can
identify two different types of modifications of the Coulomb potential $V_{C}$, 
with one of them being the standard quantum electrodynamics 
corrections, $\delta V$, satisfying 
$\left|\delta V\right|\ll\left|V_{C}\right|$
over the whole range of the radial variable $r$. The other possible
addition to $V_{C}$ is a potential arising due to the finite size of the 
atomic nucleus and as a matter of fact, can
be larger than $V_{C}$ in a very short range. We focus here on the 
latter and show that the electric potential of the proton
displays some undesirable features. Among others, the energy content
of the electric field associated with this potential is very close
to the threshold of $e^+e^-$ pair production. We contrast this large electric
field of the Maxwell theory with one emerging from the non-linear
Euler-Heisenberg theory and show how in this theory the short range
electric field becomes smaller and is well below the pair production
threshold. 
\end{abstract}


\maketitle

\section{Introduction}

The hydrogen atom was one of the stepping stones of quantum mechanics.
Today it has become a laboratory for precision physics (especially from
the experimental point of view) where several areas of physics
are combined to explain the intricacies of this rather simple system.
The inputs required are taken from particle physics \cite{bosted} 
(as such the hydrogen
atom constitutes also a testing ground for fundamental theories like
Quantum Electrodynamics (QED) and its electroweak extension \cite{testqed}), 
quantum field theories which provide
the necessary formalism (the two body Breit equation \cite{LLBook,desanctis}, 
the one-body Dirac equation or the Bethe-Salpeter equation \cite{BetheSalp}) 
and nuclear (or hadronic)
physics contributing to the physics of the finite extension of the
central nucleus \cite{ourjhepnpa,ourplb}. 
Indeed, latest efforts in the field tend to extract the static properties
of the proton from hydrogen transitions \cite{pohletc,tmart,ourjhepnpa}.  
Finally, we can imagine the hydrogen atom as a testing ground for
new ideas. From the theoretical point of view, the electroweak corrections
to the Coulomb potential $V_{C}$ play a crucial role in understanding
the simple two body system of the hydrogen atom. The high precision
of experimental confirmation of such corrections underpins the Standard
Model of particle physics or, in case of disagreement, reveals new
physics. These electroweak corrections comprised in $\delta V$ satisfy
usually the inequality $\left|\delta V\right|\ll\left|V_{C}\right|$
over the full range of the radial variable $r$. Life would be relatively
easy up to this point, were it not for hadronic corrections due to
the finite size of the proton. This second class of corrections lacks
the high precision of the electroweak corrections and therefore blurs
the extraction of the latter. Such a situation is also encountered
in other areas of precision physics (see, e.g., the problem of the
anomalous magnetic moment of the muon \cite{jeger}). 
One can circumvent
the problem of the hadronic uncertainties by trying to determine experimentally
the finite size corrections (FSC) relying on the correctness and accuracy
of the other corrections. For instance, in its simplest version, 
the FSC to the ground states can be parametrized by the
proton charge radius $<r^{2}>=\int\rho_p(r)\,r^{2}d^{3}r$ where $\rho_p(r)$
is the charge density of the proton. The  
energy correction (also in its simplest form) is given by 
$\delta E=(e^{2}/6)|\Psi(0)|^{2}<r^{2}>$ \cite{itzyk}. 
Comparing the transition energies, $\Delta E^{exp}$, with the theoretical 
ones given as $\Delta E^{\rm point}(theory) \, + \delta E^{FSC}$, 
one can extract the proton radius from
the measured transition energies. By using one static
property of the proton (the radius), one of course loses some insight
into the full structure of the extended proton. Indeed, the finite proton size 
is characterized by all moments and not only the second one
which is the radius. The charge distribution $\rho_p$ determines 
the electric field $\mathbf{E}_{p}$, the electric potential $V_{p}$ 
and partly the scalar interaction
potential $V_{ep}$ between the electron and the proton. However, by extracting  
$<r^{2}>$ alone, we do not obtain the full information of the electromagnetic
structure of the proton. In the present work we show 
that it is worthwhile 
to consider the role of the full proton structure in the hydrogen atom 
as it exhibits novel and interesting features. 
First of all, $V_{p}$ is not a correction
to the Coulomb potential in the short range. At the same time the
fact that the deviation is of a short range character allows us to treat
it as a correction in calculating corrections to the hydrogen energies
by means of time-independent perturbation theory. Secondly, as we
will show, the energy content of the field $E_{p}$ is very close to the pair
production threshold. Given the uncertainties in the electromagnetic
form factors on which $E_{p}$ is based this is a rather disturbing
fact.

Finally, the scalar interaction potential $V_{ep}$ which includes
now the electron Darwin term \cite{ourjhepnpa} 
has a large repulsive core close to the
center. By reversing the sign of the electron charge we arrive at
the interaction Hamiltonian of the positron-proton system. The repulsive
core becomes a deep potential well. 
The consequence of such a well
is a possible bound state or resonance not observed in any experiment.
Taking all this together, it appears that the proton electric field
inside the hydrogen atom might be too strong to be realistic. Of course,
even though the parametrizations of the electromagnetic form factors
lack a high precision, there is no doubt on the correct magnitude of
these form factors or on the method to calculate $E_{p}$ by using
them within the framework of the Maxwell's equations. If we want to
lower $E_{p}$ we will have to go one step ahead and modify the Maxwell's 
equations. Such a modification cannot be arbitrary, but must be a
consequence of a deeper principle. Indeed, it is well known that quantum
mechanics modifies the Maxwell's equations by the existence of a four-photon
vertex due to light-light scattering. This leads us to the 
Euler-Heisenberg theory, which is a nonlinear version of electrodynamics 
\cite{EulerH}.
In its electrostatic limit it is possible to calculate the modified
electric field $E_{p}^{\gamma\gamma}$ once the Maxwellian result
$E_{p}$ is given, i.e. we have $E_{p}^{\gamma\gamma}[E_{p}]$. One
can show that $E_{p}^{\gamma\gamma}$ is a correction to $E_{p}$
for large $r$, but in the short range region the deviations are more
significant \cite{MandAPRA,MNandAAnn}. 
Indeed, $E_{p}^{\gamma\gamma}$ turns out to be much smaller
than $E_{p}$ in this region which, in principle, is the effect we
are looking for. There is a priori no reason to discard this short
range region as its only observable manifestation will be in the
energy corrections which will be small due to the short range character
(the situation is similar to the case of $E_{p}$ in comparison to
the Coulomb potential). However, the magnitude of the field will exceed
the limits of validity of the Euler-Heisenberg theory valid for weak
fields. This can be traced back partly to the large values of $E_{p}$
close to the center. We will show that the inclusion of higher orders
improves the situation. We speculate that already a third order result
might be sufficient to reduce the Maxwellian result of $E_{p}$ and
at the same time obeying the weak field condition imposed in the 
Euler-Heisenberg
theory. In the second step we show how the light-light formalism can
be extended to the whole scalar interaction in the hydrogen atom,
this is to say including the electron Darwin term.

The notation convention in the present work consistently follows \cite{LLBook}. This 
implies, $e^2 = \alpha$ and the Breit equation as well as the Euler-Heisenberg 
Lagrangian are as in \cite{LLBook}.

The article is organized as follows. In Section II we discuss
the derivation of the scalar interaction potential in the hydrogen
atom via the Breit equation with form factors. We identify here also the part which
is due to the electric potential of the proton and compare it with the point-like 
Coulomb result.
In the Section III, 
we give an overview of the Euler-Heisenberg theory, specializing on
the electrostatic case. In the fourth section we apply the results
of the Euler-Heisenberg theory to the hydrogen atom giving a new perspective
to the short range interaction. 
In Section V, we comment on the truncation of the expansion which we use in the 
article. 
In the last section we draw our conclusions.

\section{Breit equation with form factors}

In accordance with scattering theory in quantum mechanics, 
the Born approximation of the scattering amplitude is proportional to 
the Fourier transform of the potential. Vice versa, given a scattering
amplitude from the Feynman diagrams suitably expanded in powers of $1/c$
($c$ is the velocity of light) we can derive a potential in $\mathbf{r}$
space. This well known procedure has been applied many times in physics
to get some insight into old and new dynamics. One of the application
of this principle is the Breit equation which starting from the elastic
electron proton scattering amplitude gives us some well known
interaction terms (like the Coulomb, electron Darwin, fine and hyperfine
structure \cite{fabian}) and some new input in the non-relativistic picture (like
the proton Darwin and retardation terms \cite{fabian,ourjhepnpa}). 
Finally, it is possible to include in the Breit equation, the finite size 
corrections in an elegant
way by using the most general current of the proton and replacing 
\begin{equation}
e\gamma_{\mu}\to e\left(F_{1}(q^{2})\gamma_{\mu}+\frac{F_{2}(q^{2})}{2m_{p}}\sigma_{\mu\nu}q^{\nu}\right)\label{replacing}
\end{equation}
In case of a point-like object, $F_{1}=1$ and $F_{2}=0$. 
The following Breit equation is thus quite general: 

\begin{eqnarray}
 &  & \hat{U}(\textbf{p}_{X},\textbf{p}_{p},\textbf{q})=4\pi e^{2}\Bigg[F_{1}^{X}F_{1}^{p}\Bigg(-\frac{1}{\textbf{q}^{2}}+\frac{1}{8m_{X}^{2}c^{2}}+\frac{1}{8m_{p}^{2}c^{2}}+\frac{i\bm{\sigma}_{p}.(\textbf{q}\times\textbf{p}_{p})}{4m_{p}^{2}c^{2}\textbf{q}^{2}}-\frac{i\bm{\sigma}_{X}.(\textbf{q}\times\textbf{p}_{X})}{4m_{X}^{2}c^{2}\textbf{q}^{2}}\nonumber \\
\nonumber \\
 &  & +\frac{\textbf{p}_{X}.\textbf{p}_{p}}{m_{X}m_{p}c^{2}\textbf{q}^{2}}-\frac{(\textbf{p}_{X}.\textbf{q})(\textbf{p}_{p}.\textbf{q})}{m_{X}m_{p}c^{2}\textbf{q}^{4}}-\frac{i\bm{\sigma}_{p}.(\textbf{q}\times\textbf{p}_{X})}{2m_{X}m_{p}c^{2}\textbf{q}^{2}}+\frac{i\bm{\sigma}_{X}.(\textbf{q}\times\textbf{p}_{p})}{2m_{X}m_{p}c^{2}\textbf{q}^{2}}+\frac{\bm{\sigma}_{X}.\bm{\sigma}_{p}}{4m_{X}m_{p}c^{2}}\nonumber \\
\nonumber \\
 &  & -\frac{(\bm{\sigma}_{X}.\textbf{q})(\bm{\sigma}_{p}.\textbf{q})}{4m_{X}m_{p}c^{2}\textbf{q}^{2}}\Bigg)+F_{1}^{X}F_{2}^{p}\Bigg(\frac{1}{4m_{p}^{2}c^{2}}+\frac{i\bm{\sigma}_{p}.(\textbf{q}\times\textbf{p}_{p})}{2m_{p}^{2}c^{2}\textbf{q}^{2}}-\frac{i\bm{\sigma}_{p}.(\textbf{q}\times\textbf{p}_{X})}{2m_{X}m_{p}c^{2}\textbf{q}^{2}}-\frac{(\bm{\sigma}_{X}.\textbf{q})(\bm{\sigma}_{p}.\textbf{q})}{4m_{X}m_{p}c^{2}\textbf{q}^{2}}\nonumber \\
\nonumber \\
 &  & +\frac{\bm{\sigma}_{X}.\bm{\sigma}_{p}}{4m_{X}m_{p}c^{2}}\Bigg)+F_{2}^{X}F_{1}^{p}\Bigg(\frac{1}{4m_{X}^{2}c^{2}}-\frac{i\bm{\sigma}_{X}.(\textbf{q}\times\textbf{p}_{X})}{2m_{X}^{2}c^{2}\textbf{q}^{2}}+\frac{i\bm{\sigma}_{X}.(\textbf{q}\times\textbf{p}_{p})}{2m_{X}m_{p}c^{2}\textbf{q}^{2}}-\frac{(\bm{\sigma}_{X}.\textbf{q})(\bm{\sigma}_{p}.\textbf{q})}{4m_{X}m_{p}c^{2}\textbf{q}^{2}}\nonumber \\
\nonumber \\
 &  & +\frac{\bm{\sigma}_{X}.\bm{\sigma}_{p}}{4m_{X}m_{p}c^{2}}\Bigg)+F_{2}^{X}F_{2}^{p}\Bigg(\frac{\bm{\sigma}_{X}.\bm{\sigma}_{p}}{4m_{X}m_{p}c^{2}}-\frac{(\bm{\sigma}_{X}.\textbf{q})(\bm{\sigma}_{p}.\textbf{q})}{4m_{X}m_{p}c^{2}\textbf{q}^{2}}\Bigg)\Bigg],\label{potFF}
\end{eqnarray}
where $X$ stands for electron or muon. In the rest of the article we
will focus on the scalar part of this equation neglecting the structure
of the point-like lepton. This implies that we will not take into account
terms with spins or momentum dependent interactions. We are then left
with the $ep$ interaction potential given by \cite{ourjhepnpa},  
\begin{equation}
V_{ep}({\bm{q}})=-4 \pi e^2\left[\frac{\tilde{G}_{E}({\bm{q}})}{{\bm{q}}^{2}}-\frac{1}{8m_{p}^{2}c^{2}}\tilde{G}_{E}({\bm{q}})-\frac{1}{8m_{e}^{2}c^{2}}\tilde{G}_{E}({\bm{q}})
\right]\label{GE}\, ,
\end{equation}
where $\tilde{G}_{E}({\bm {q}} = G_{E}({\bm {q}})(1 - {\bm {q}}^2/8m_p^2)$. 
The above, in principle, is valid at higher order in the $1/c$ expansion.
At the lowest order of the non-relativistic expansion we have 
\begin{eqnarray}
\tilde{G}_{E}\simeq G_{E} & = & F_{1}-\frac{\mathbf{q}^{2}}{4m_{p}^{2}c^{2}}F_{2}\nonumber \\
\tilde{G}_{M}\simeq G_{E} & = & F_{1}+F_{2}
\end{eqnarray}
where $G_{E}$ and $G_{M}$ are the electric and magnetic Sachs form
factors. With this approximation, the first term in 
the square brackets in (\ref{GE}), called
the Coulomb term, will become $G_{E}/\mathbf{q}^{2}$ whereas the
electron Darwin term is well approximated by $V_{eD}=4 \pi e^2 (1/8m_{e}^{2}c^{2})F_{1}$
(and similarly the proton Darwin term $V_{pD}=4 \pi e^2 (1/8m_{p}^{2}c^{2})F_{1}$).
The Fourier transform of (\ref{GE}) gives 
\begin{eqnarray}
V_{ep}({\bm{r}}) & = & \int\frac{d^{3}q}{(2\pi)^{3}}e^{i{\bm{q}}\cdot{\bm{r}}}V_{ep}({\bm{q}})\nonumber \\
 & = & -e[V_{C}({\bm{r}})+V_{pD}({\bm{r}})+V_{eD}({\bm{r}})]= - e[V_{p}({\bm{r}})+V_{eD}({\bm{r}})]
\end{eqnarray}
and we can identify the electric potential of the proton $V_{p}$ from
the Breit equation to be given by the parts which, apart from being scalar 
(no spin and momentum dependence) are also independent of the properties of 
the probe (in this case the lepton).
By this token the proton Darwin term, $V_{pD}$, is part of the electric potential,
but the electron Darwin, $V_{eD}$, not. This has some consequences for the electric
field at small distances inside the proton and also for the charge
distribution, but is of no further importance here. Using the Poisson
equation we can also say more about both the Darwin terms in $\mathbf{r}$
space, i.e., we have 
\begin{equation}
V_{eD}({\bm{r}})\propto V_{pD}({\bm{r}})\propto\rho({\bm{r}})\propto\int d^{3}qe^{i{\bm{q}}\cdot{\bm{r}}}\tilde{G}_{E}({\bm{q}})\label{Darwinrho}
\end{equation}
Expanding the Dirac Hamiltonian (Ref. \cite{LLBook}, p.125) up
to order $1/c^{2}$ gives the well known result 
\begin{equation}
\hat{H}_{Dirac}\simeq\frac{\hat{\mathbf{p}}^{2}}{2m_{e}}-\frac{\hat{\mathbf{p}}^{4}}{4m_{e}^{2}c^{2}}+eV_{p}-\frac{e}{4m_{e}^{2}c^{2}}\mathbf{\sigma}\cdot\mathbf{E}\times\hat{\mathbf{p}}-\frac{e}{8m_{e}^{2}c^{2}}\nabla\cdot\mathbf{E} \, . \label{Dirac}
\end{equation}
The last term is the electron Darwin term which using the Maxwell's 
equations is, of course, proportional to $\rho$. 
If we go from the Maxwell theory to the Euler-Heisenberg
theory, the Gauss law is not valid anymore and it therefore matters
if we treat the electron Darwin term in the Breit equation formalism 
or the Dirac one. 
Apart from this, we note that the fact that no
proton Darwin term appears in the Dirac equation is based on the difference
between the Breit and the Dirac equation. The Breit equation is a
true two-body equation whereas the Dirac equation is a one particle
equation. We will come back to this point later. 

\subsection{Scalar potentials of the extended proton}

It is sometimes useful to resort to simplifying assumptions in order to obtain 
analytical formulae providing a physical insight into the problem, rather than 
using more precise inputs which do not lead to analytical expressions and are difficult
to handle numerically. 
Take, for instance, the energy
correction $\delta E$ due to the finite size effects in the hydrogen
atom. The calculation is, in principle, a two scale problem where one 
encounters the nuclear form factors which are important at distances of only a  
couple of fermis combined with the atomic wave function which extends 
to about 10$^4$ fm. The combination of scales makes a numerical calculation 
difficult and hence it is convenient to use the dipole parametrization for 
the proton form factor which allows analytical calculations. Hence we choose, 
\begin{equation}
G_{E}({\bm{q}}^{2})=1/(1+{\bm{q}}^{2}/m^{2})^{2} \, .\label{dipole}
\end{equation}
This leads to the electric potentials, 
\begin{eqnarray}\label{electricpots}
V_{p}(r) & = & V_{C}+V_{pD}\, ,\,\,{\rm with}\nonumber \\
V_{C}(r) & = & \frac{e}{r}\left(1-e^{-mr}\left(1+\frac{mr}{2}\right)\right)\nonumber \\
V_{pD}(r) & = & -\frac{e}{8m_{p}^{2}}\left[\left(1+ \frac{\kappa_p(1-2\kappa^2)}{1-\kappa^2}\right)\frac{m^{3}}{2}e^{-mr}
-\frac{\kappa^{2}\kappa_{p}m^{2}}{(1-\kappa^{2})^{2}}\frac{e^{-mr}}{r} +\frac{\kappa^{2}\kappa_{p}m^{2}}{(1-\kappa^{2})^{2}}\frac{e^{-\kappa mr}}{r}\right] \, ,
\end{eqnarray}
where $\kappa=2m_{p}/m$ and $\kappa_{p}$ the anomalous magnetic moment
of the proton. 
To arrive at the interaction potential it suffices to multiply the
above expressions with $-e$ and add the electron Darwin term. 
The main conclusions of the paper will
remain mostly insensitive to the parametrization of the electric form-factors
as we will explain below.

\subsubsection{$ep$ Interaction Potentials} 
Let us have an unbiased look at the interaction potentials
depicted in Figures 1-3.  
\begin{figure}[h]
\includegraphics[width=6cm,height=6cm]{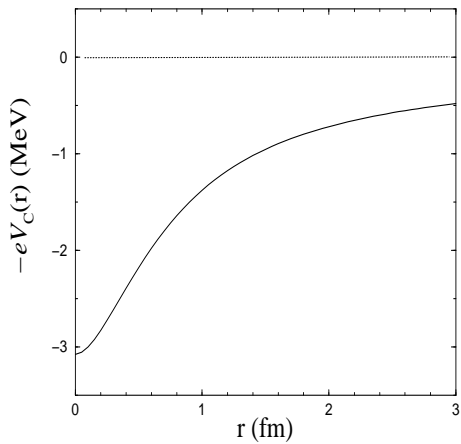}
\caption{The electric short range interaction potential $-eV_{C}(r)$
of the proton with the electron (only Coulomb with finite size and 
no Darwin terms included). }
\end{figure}

\begin{figure}[h]
\includegraphics[width=8cm,height=7cm]{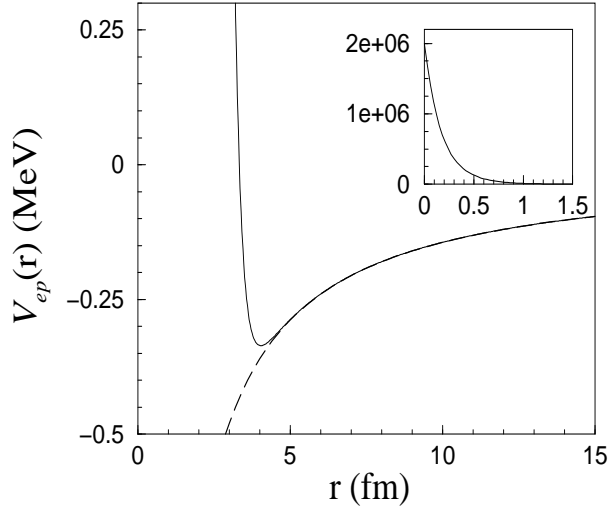}
\caption{The short range $ep$ interaction potential including
the electron Darwin term. The dashed line represents the point-like
Coulomb potential $-\alpha/r$. The short distance behavior is mostly
due to the electron Darwin term. This is shown in the inlay.}
\end{figure}

\begin{figure}[h]
\includegraphics[width=8cm,height=8cm]{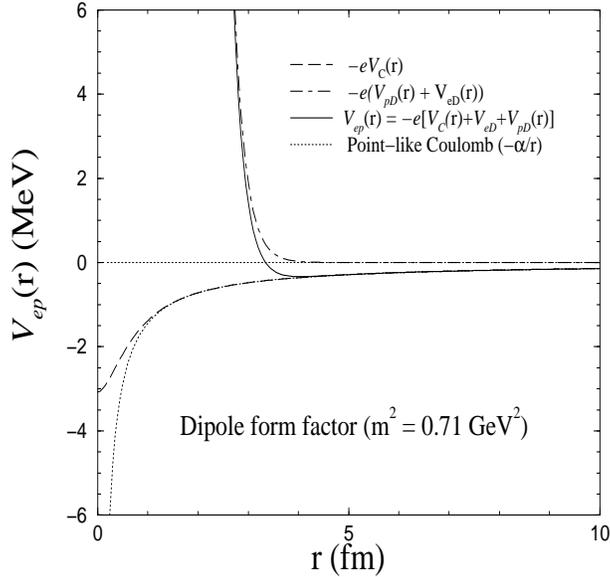}
\caption{Comparison of the different finite size corrections to the
electron-proton interaction potential.}
\end{figure}

The first one in Fig. 1 resembles a Wood-Saxon potential with a depth of $-3$
MeV. The second one (Fig. 2) has a minimum at $-0.3$ MeV, but also a repulsive
core with a height of $2\times10^{3}$ MeV. The potentials are, of course,
not nuclear potentials (although they have to do with hardonic properties),
but reflect the finite size of the proton. It is interesting to infer
on the effect of these potentials from an angle which is different from the 
time independent perturbation theory. To this end we choose the variational
principle (see next sub-section) which reveals that the ground state
is close to $-13$ eV. In spite of the modifications of the electron
proton potential at short distances the resulting energy eigenvalue
is still like in the standard hydrogen atom (plus corrections,
of course). On the other hand, by replacing in $V_{ep}$, $e\to-e$, 
one obtains the positron proton potential with a potential depth of
$10^{6}$ MeV. The question we can ask here is if such a deep but short
range potential will lead to a bound state or a resonance.

\subsubsection{Proton electric potential} 
Before proceeding, let us address two issues connected with the electric potential
of the proton. The first one is the inclusion of the proton Darwin
term into the definition of the electric potential. Apart from being
suppressed by the squared proton mass, the effects are mild as shown
in Fig. 4 for the charge distribution (evaluated using 
$\nabla^2 V_p(r)$). 
However, this term is 
large at short distances and we shall come to the effects of it later. 
\begin{figure}[h]
\includegraphics[width=8cm,height=8cm]{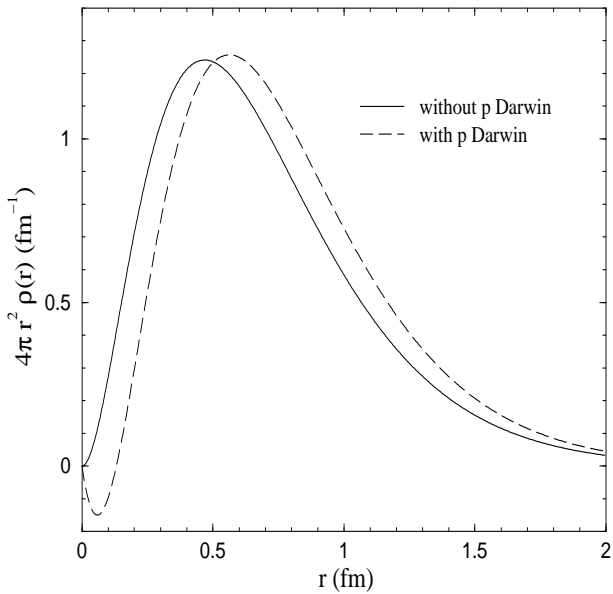}
\caption{The charge distribution of the proton.}
\end{figure}
Due to ${\bm{\nabla}}\cdot{\bm{E}}=\frac{2}{r}E+\frac{dE}{dr}\propto\rho$, 
the charge distribution is non-zero at the center when we include
the proton Darwin term in the potential. The second issue involves testing 
the sensitivity of the results to the choice of the parametrization of the
electromagnetic form factors. This is
shown in Fig. 5. The main observation of the large value of the short range
potentials does not change qualitatively for the three parametrizations.
Notice also that all three cases display a local minimum albeit displaced by
one fermi. This displacement is not of much significance for our main points. 
Therefore, we continue working with the dipole parametrization in the rest of the work. 

\begin{figure}[h]
\includegraphics[width=8cm,height=8cm]{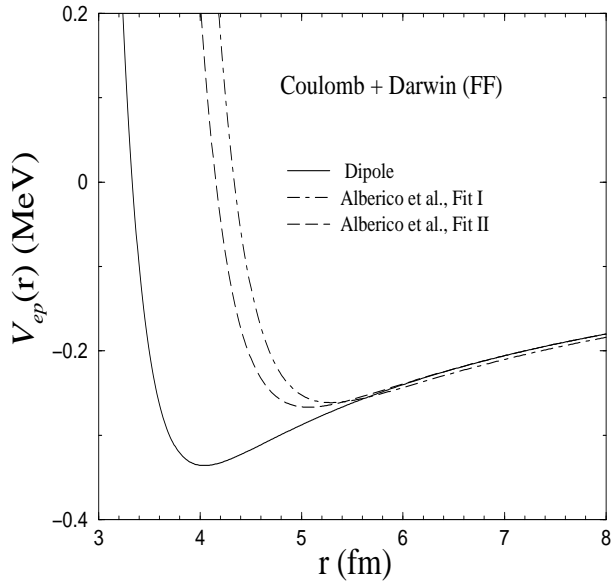}
\caption{The electron-proton interaction potential in different 
parametrizations.}
\end{figure}

It is not common to see results similar to those presented here (and
in the subsequent sub-section) very often since the simplest estimate
of the correction to the energy level(s) of the hydrogen atom is 
\begin{eqnarray}
\delta{\cal E} & = & \int d^{3}x|\Psi(x)|^{2}\left[eV_{p}(r)+\frac{e^{2}}{r}\right]\nonumber \\
 & \simeq & |\Psi(0)|^{2}\int d^{3}x\left[eV_{p}(r)+\frac{e^{2}}{r}\right]
\end{eqnarray}
which after integration by parts becomes 
\begin{equation}
\frac{e}{6}|\Psi(0)|^{2}\int d^{3}xr^{2}{\bm{\nabla}}^{2}V_{p}\label{corr2}\, .
\end{equation}
Using the Poisson equation and the definition of the radius as the second
moment of the charge distribution, i.e., 
\begin{equation}
<r^{2}>=\int d^{3}x\rho r^{2}\label{corr3}\, , 
\end{equation}
one finally gets 
\begin{equation}
\delta{\cal E}=\frac{e}{6}|\Psi(0)|^{2}Ze<r^{2}>\label{corr4}\, . 
\end{equation}
Seemingly one does not need to know anything about the electric fields
inside the proton as everything is encoded in the proton radius. 
However, a calculation of the electric field 
corresponding to this potential (which will be 
presented in the next section) shows that it is extremely large towards the center. 
One can calculate the energy content in
the electric field:  
\begin{equation}
\mathscr{E}[E_{p}]=(1/8\pi)\int E_{p}^{2}d^{3}x\simeq1\,\,\, MeV\, ,\label{corr5}
\end{equation}
which is very close to $2m_{e}$ and given the uncertainty of the
form-factors at the border of the $e^+e^-$ pair production threshold.
One is justified to ask if the field inside the hydrogen atom can be so large 
and dangerously close to the pair production threshold. Finally, is there 
a way to reduce it?

\subsubsection{Some important perspectives}
Before embarking on the variational calculations it is illuminating
to give the finite size corrections or, in other words the potential
at short distances, some other perspectives. Even at the cost that this is 
obvious let us remark that the field/potential
of the proton at small $r$ is large, but still finite as compared
to $1/r$. Therefore, it is not a correction, but its short range
allows us to treat it as a correction in $\delta{\cal E}$. We shall discuss this 
further in the next subsection.

Classically, by Gauss law, the electron will not ``feel'' the finite
size of the proton (assuming its charge distribution to be spherically
symmetric). Hence, the fact that we can calculate it and measure it
is indeed a quantum effect which would not be present in the classical
treatment as long as we treat the proton also classically, i.e., as
an object with a sharp radius (hard sphere). The classical electron
will only see the Coulomb potential as long as it is outside the proton.
If we relax the classical picture and replace the sharp proton by
a ``medium'' with a charge distribution without border then the
classical electron will also get affected by the $V_{ep}$ depicted
in figures 1-3. 
For example, let us consider the potential (solid line) in Fig. 2 with 
a depth of around 350 keV. 
In this case, a classical ground state which is a few hundred keV deep 
and with an orbit at a radius of about 4 fm is possible. 
Classically, the electron in spite of
being accelerated cannot radiate losing energy as $E=V_{ep}$ ($l=0$)
is classically the lowest possible energy. This means that the local
minimum around 350 keV would be the real minimum of the electron in 
the hydrogen atom. 
A quantum mechanical calculation with the same potential, however, reproduces 
the realistic picture of a 
ground state close to 13.6 eV and with a radius which is about 4 orders 
of magnitude larger and in agreement with the atomic radius. The effect of the 
modified potential (as compared to $1/r$) is only a tiny correction to the 
ground state energy. 

The quantum effect mentioned above is based on the fact that we need
to define the wave function everywhere in space and as a result also
the potential. Hence, it is expected that we will find more examples
in nature manifesting such a quantum effect. Indeed, an electron around
a spherically symmetric mass distribution (formally, this has to do
with the Dirac equation in the Schwarzschild metric) will know if
it ``moves'' around a black hole or a star \cite{blackholeelectron}. It can be
shown that in the case of a black hole there are no bound states which is 
partly a macroscopic quantum effect related to the same fact that quantum
mechanically the electron ``knows'' that the proton is extended.

\subsection{Variational approach to finite size corrections}

The usual separation of variables in the Schr\"odinger equation gives
the following equation for the radial function $R(r)$,  
\begin{equation}
-\frac{1}{2\mu}\left(\frac{2}{r}\frac{dR}{dr}+\frac{d^{2}R}{dr^{2}}\right)+
V(r)R=ER \, ,\label{E1}
\end{equation}
which, using the definition: $R=\chi(r)/{r}$, 
becomes 
\begin{equation}
\frac{-1}{2\mu}\frac{d^{2}\chi}{dr^{2}}+V(r)\chi=E\chi.
\end{equation}
with the normalization condition $\int_{0}^{\infty}dr\chi^{2}(r)=1$.

We denote the expectation value of the kinetic energy operator $\frac{p^{2}}{2m_{e}}=-\frac{\nabla^{2}}{2m_{e}}$
by $E_{{\rm kin}}$ which in terms of $\chi$ is given as 
\begin{equation}
E_{{\rm kin}}=\frac{1}{2m_{e}}\int_{0}^{\infty}dr\left(\frac{d\chi}{dr}\right)^{2}.\label{kin} 
\end{equation}
With the inclusion of both the Darwin terms, it is convenient to write
the potential as 
\begin{equation}
V_{ep}(r)=\frac{-\alpha}{r}+C_{1}e^{-mr}+C_{2}\frac{e^{-mr}}{r}+C_{3}\frac{e^{-mkr}}{r}\label{E6}
\end{equation}
with 
\begin{eqnarray}
C_{1} & = & \alpha\left[A\frac{m^{3}}{2}+A\frac{m^{3}\kappa_{p}}{2(1-\kappa^{2})}+\frac{m}{2}\right]\nonumber \\
C_{2} & = & \alpha\left[1 + A\frac{m^{2}\kappa^{2}\kappa_{p}^{2}}{(1-\kappa^{2})^{2}}\right]\nonumber \\
C_{3} & = & -C_{2}+ \alpha 
\end{eqnarray}
with $A=1/(8m_{p}^{2})+1/(8m_{e}^{2})$, $\kappa=\frac{2m_{p}}{m}$
and $m^{2}=0.71\,\,{\rm GeV}^{2}$ is the parameter from the
dipole parametrization.

All possible trial radial wave functions are subjected to the following
condition resulting from the Schr\"odinger equation at the origin: 
\begin{equation}
-\frac{1}{\mu r}\left(R'(0)+rR''(0)+\ldots\right)-\frac{1}{2m_{e}}R''(0)+R(0)\lim_{r\rightarrow0}V(r)=ER(0).\label{E2}
\end{equation}
Since we have $C_{2}+C_{3}=\alpha$, the condition which ensures the
regularity at the origin is simply $R'(0)=0$. One trial wave function
which satisfies this condition is 
\begin{equation}
R(r)=N(e^{-\eta r}-\frac{1}{d}e^{-d\eta r})\, ,\label{trial1}
\end{equation}
where $d$ is a parameter which we choose later to be 2. The normalization
constant is easily calculated to be 
\begin{equation}
N=\sqrt{\frac{4 \eta^{3}}{\left(\frac{1}{d^{5}}-\frac{16}{(d+1)^{3}d}+1\right)}}\, . 
\end{equation}
The kinetic part of the expectation value of energy is 
then $E_{k}=\frac{1}{2m_{e}}\frac{92\eta^{2}}{127}$.
The potential term is obtained from the expression 
$E_{{\rm pot}}=\int_{0}^{\infty}drV(r)R^{2}(r)r^{2}$
which gives,  
\begin{eqnarray}
E_{pot} & = & -\alpha N^{2}\left(\frac{1}{(2\eta)^{2}}-\frac{1}{(3\eta)^{2}}+\frac{1}{4(4\eta)^{2}}\right)\nonumber \\
 & + &  C_{1}N^{2}\left(\frac{2}{(2\eta+m)^{3}}-\frac{2}{(3\eta+m)^{3}}+\frac{2}{4(4\eta+m)^{3}}\right)\nonumber \\
 & + &  C_{2}N^{2}\left(\frac{1}{(2\eta+m)^{2}}-\frac{1}{(3\eta+m)^{2}}+\frac{1}{4(4\eta+m)^{2}}\right)\nonumber \\
 & + &  C_{3}N^{2}\left(\frac{1}{(2\eta+km)^{2}}-\frac{1}{(3\eta+km)^{2}}+\frac{1}{4(4\eta+km)^{2}}\right).
\end{eqnarray}
Minimizing $E(\eta)=E_{{\rm pot}}(\eta)+E_{{\rm kin}}(\eta)$ with
respect to the variational parameter $\eta$ we find $\eta=0.00432\,\,{\rm MeV}$.
This corresponds to $E=-13.274\,\,{\rm eV}$ and $\sqrt{<r^{2}>}=0.862555\,\,{\rm fm}$
where $<r^{2}>$ is the expectation value with respect to the trial
wave function. Coming back to the large deviation of (\ref{E6}) as
compared to the Coulomb potential at short distances and noticing
the local minimum at roughly four fermi (where classically we would
find bound solutions), it is a priori not excluded to find a new bound
state close to the proton which we could be overlooked by directly applying
the perturbation method to calculate the energy correction. The variational
approach convinces us that this is not the case and we are dealing
still with the standard hydrogen atom (plus corrections). Having done
this exercise we can perform a similar (variational) calculation for
the positron-proton system by simply reversing the sign of our potential.
Again, what would be a senseless undertaking if we had only the Coulomb
potential, appears now in a different light by contemplating the deep
minimum at the center of $-V_{ep}$. For (negative) binding energies
we pay attention to the fact that the mass of the atom, i.e., $m_{e}+m_{p}+E$
must be positive. The results are extremely sensitive (the sensitivity
is not due to the variational method, but to the form of the potential)
to the choice of the electron mass. We therefore replace $m_{e}$
by $m_{x}$ and vary this parameter. We find a very narrow range of
$m_{x}$, namely $(0.48429\,{\rm MeV},0.48843\,{\rm MeV})$, where
a negative binding energy is possible. 

The second variational ansatz we use will be: 
\begin{equation}\label{ansatz2}
R(r)=Ne^{-\eta r}(1+\eta r).
\end{equation}
For this with a negative charge ($-e$) for the probe particle we
obtain for the electron mass an  
$\eta$ and for this value of $\eta$
an energy expectation value of $-13.1123$ eV. So the variational principle
seems to work fine. The RMS radius at this point is: $0.849069$ fm.\\
It is worth noting that any value of the probe mass for a charge
of $-e$ gives a bound state according to this variational principle.
The same happens for the ansatz in the preceeding
paragraphs.\\
Now for the opposite charge $+e$ for the probe particle, 
almost the same as in the variational ansatz (\ref{trial1}) occurs: there is a narrow
interval in which the energy expectation value is negative suggesting
a bound state appearance. This interval is: (0.491837 MeV,0.495918MeV).
Evidently there are approximations and inaccuracies involved
in these calculations. We took the anomalous magnetic moment of the
probe equal to zero in all cases. We use the dipole parametrization
for the form factor and another parametrization could change the 
variational calculation results as it is sensitive to the choice of the mass.

Apart from the bound state with negative binding energy
there is a clear possibility of a Gamow state for the
positron-proton case as can be seen from the potential.
This positive energy state (resembling an $\alpha$-nucleus 
case) would be located between zero and the peak of the positive bump of the
potential which the particle would traverse by the quantum mechanical tunneling
mechanism. In this case we would talk about a resonance. 

\section{Overview of the Euler-Heisenberg theory}

We give a brief account and the basic equations of the Euler-Heisenberg
Theory \cite{MandAPRA}. 
For a detailed exposition, we refer the reader to \cite{EulerH}. 
In quantum electrodynamics, the one loop exchange of electrons allows
the photon-photon interaction. Due to Furry's theorem (based on charge
conjugation invariance), only an even number of external photons are
allowed in a Feynman diagram. The Euler-Heisenberg theory considers
the $\gamma\gamma\rightarrow\gamma\gamma$ case. This self-interaction
of photons, which is not present in classical electrodynamics, is a
quantum mechanical effect that can be effectively translated into
the configuration space as an additional term in the Lagrangian. In
doing so, and under certain conditions (see below), we can use the
full Lagrangian (the classical one from Maxwell's theory and the one
resulting from the $\gamma\gamma\rightarrow\gamma\gamma$ interaction)
to construct a non-linear classical field theory. In the regimes where
applicable, this non-linear theory can be used to calculate quantum
effects that are outside the scope of the linear Maxwell's theory.

Any Lagrangian describing electromagnetic fields must be gauge and
relativistic invariant. Hence, the Lagrangian is necessarily a function
of the invariants 
\begin{eqnarray}
\textgoth{F} & \equiv & -\frac{1}{4}F_{\mu\nu}F^{\mu\nu}=\frac{1}{2}({\mathbf{E}}^{2}-{\mathbf{B}}^{2}),\label{inv}\\
\textgoth{G} & \equiv & \frac{1}{4}F_{\mu\nu}\tilde{F}^{\mu\nu}={\mathbf{E}}\cdot{\mathbf{B}}.\nonumber 
\end{eqnarray}

Assuming constant and uniform fields, the one loop effective Lagrangian
of QED encoding the light-light interaction is \cite{Euler-Heisenberg, Schwinger}  

\begin{equation}
{\cal L}={\cal L}_{0}+{\cal L}_{EH},\label{L1}
\end{equation}
where the Maxwell's Lagrangian is

\begin{equation}
{\cal L}_{0}= \frac{1}{4\pi} \,\textgoth{F} \label{L2-1}
\end{equation}
and the Euler-Heisenberg term is 
\begin{eqnarray}
\mathcal{L}_{EH} & = & -\frac{1}{8\pi^{2}}\int_{0}^{\infty}ds\, s^{-3}e^{-m_{e}s}\nonumber \\
 &  & \times\left[\left(es\right)^{2}(ab)\coth\left(esb\right)\cot\left(esa\right)-1-\frac{2}{3}\left(es\right)^{2}\textgoth{F}\right],\label{EH83}
\end{eqnarray}
 with
$a = |e| E/m_e^2$ and $b = |e| B/m_e^2$. 
For weak fields the Lagrangian (\ref{EH83}) can be expanded as an
asymptotic series \cite{weakfieldL}. 
The first term in the expansion correcting the Maxwell's Lagrangian is 
\begin{equation}
{\cal L}_{{\rm EH}}^{(1)}=\eta\left(({\mathbf{E}}^{2}-{\mathbf{B}}^{2})^{2}+7({\mathbf{E}}\cdot{\mathbf{B}})^{2}\right)\label{L3}
\end{equation}
with 
\begin{equation}
\eta\equiv\frac{\alpha^{2}}{360\pi^{2}m_{e}^{4}},\label{L33}
\end{equation}
where we have set $\hbar=c=1$. 

As previously mentioned, the new Lagrangian can be used to construct
a classical theory of fields. The Lagrangian (\ref{L3}) gives rise
to new Maxwell's equations (for the moment we will restrict ourselves
to the lowest order in the fine structure constant $\alpha$ and discuss
higher orders below) which due to the self-interaction resemble the
standard Maxwell equation in matter \cite{LLBook}. In this theory,
the homogeneous Maxwell's equations remain unchanged (since they define
the electromagnetic potentials): 
\begin{eqnarray}
{\mathbf{\nabla}}\cdot{\mathbf{B}} & = & 0,\label{max1}\\
{\mathbf{\nabla}}\times{\mathbf{E}}+\frac{\partial{\mathbf{B}}}{\partial t} & = & 0.\nonumber 
\end{eqnarray}
The inhomogeneous new Maxwell's equations follow from a variation of
the Euler-Heisenberg Lagrangian. Let us define four auxiliary vectors:
the displacement and the polarization vector 
\begin{eqnarray}
{\mathbf{D}} & \equiv & {\mathbf{E}}+4\pi{\mathbf{P}},\label{max2}\\
{\mathbf{P}} & \equiv\frac{\partial{\cal L}_{{\rm EH}}^{(1)}}{\partial\mathbf{E}}= & \eta\left[4{\mathbf{E}}({\mathbf{E}}^{2}-{\mathbf{B}}^{2})+14{\mathbf{B}}({\mathbf{E}}\cdot{\mathbf{B}})\right];\nonumber 
\end{eqnarray}
and the magnetic intensity and the magnetization vector 
\begin{eqnarray}
{\mathbf{H}} & \equiv & {\mathbf{B}}-4\pi{\mathbf{M}},\label{max2.2}\\
{\bm{M}} & \equiv & \frac{\partial{\cal L}_{{\rm EH}}^{(1)}}{\partial\mathbf{B}}=\eta\left[4{\mathbf{B}}({\mathbf{E}}^{2}-{\mathbf{B}}^{2})-14{\mathbf{E}}({\mathbf{E}}\cdot{\mathbf{B}})\right].\nonumber 
\end{eqnarray}
In terms of these vectors, the new inhomogeneous Maxwell's equations
read 
\begin{eqnarray}
{\mathbf{\nabla}}\cdot{\mathbf{D}}=\rho,\label{max3}\\
{\mathbf{\nabla}}\times{\mathbf{H}}-\frac{\partial\mathbf{D}}{\partial t}={\mathbf{j}}.\nonumber 
\end{eqnarray}

As in classical electrodynamics, given the charge density and the
current distribution, the electric and magnetic fields can be calculated
by solving the new Maxwell's equations.

We take the point of view that the external sources $\rho$ and $\mathbf{j}$
are the same whether we consider the Maxwell or the Euler-Heisenberg
theory. Of course, the resulting fields will differ. We mention that
this natural point of view leaves untouched many standard definitions
and results. For instance, the charge radius of the proton (or any
other particle), defined by 
\begin{equation}
<r^{2}>=\int d^{3}x\, r^{2}\rho,\label{radius}
\end{equation}
remains the same in both cases.

\subsection{Ranges of validity of the theory}
The Euler-Heisenberg theory is an approximation. In order for the
approximation to be valid, the fields have to obey certain conditions.
Firstly, the electric field should never surpass the so called critical
field 
\begin{eqnarray}\label{criticalf}
E & \ll & E_{c}\equiv\frac{m_{e}^{2}}{\sqrt{\alpha}}\approx 3\,{\rm MeV^{2}}.\label{Ec}
\end{eqnarray}
The reason for this requirement is to keep small the probability of the
electron pair production 
\begin{equation}
P_{e^{+}e^{-}}\propto\exp\left(-\frac{\pi m_{e}^{2}}{|e|E}\right),\label{pair}
\end{equation}
 as it should be if we want to describe the electromagnetic interaction
using classical fields.
If we wish to use the weak field expansion, then the magnetic field
has to obey a similar condition: 
\begin{equation}
B\ll B_{c}\equiv\frac{m_{e}^{2}}{\sqrt{\alpha}}.
\end{equation}

Secondly, as mentioned in the previous section, the Euler-Heisenberg
Lagrangian is derived assuming the fields to be uniform in the whole
space. The energy associated with such fields is then infinite. However,
as long as the fields obey
\begin{align}\label{conditions}
\frac{1}{m_{e}}\left|\nabla E\right| & \ll E,\\
\frac{1}{m_{e}}\left|\nabla B\right| & \ll B,
\end{align}
the Euler-Heisenberg theory should give good results when applied
to non-uniform fields.

A word of caution is in order here. For a Coulomb field, we obtain 
from (\ref{conditions}) a restriction for the range of applicability of the 
Euler-Heisenberg theory in the form, $r >$ 772 fm. Taken at face 
value it would prevent us from using the Euler-Heisenberg theory for fields 
inside the proton. However, we note that (i) the numerical restriction is derived 
from a Coulomb field, whereas the electric field inside the proton is not 
Coulomb-like (due to form factors as can be seen in Fig. 7) 
and (ii) we assume that it is sufficient that Eqs (\ref{criticalf}) and 
(\ref{conditions}) are satisfied after the field is calculated from the 
Euler-Heisenberg theory. Furthermore, we mention that (\ref{conditions}) is 
satisfied in a small region around 0.34 fm when the electric field is evaluated 
from $V_C(r)$ given in (\ref{electricpots}). 
We shall come back to this point later.  

A necessary condition for pair production in non-uniform fields is
that the energy content of the fields has to be large enough to reach 
the mass of the particles that could be produced. The question
which arises is if we should consider the pair production threshold
before or after the Euler-Heisenberg corrections to the fields. This
is a crucial point since we will show that the Euler-Heisenberg corrections
can reduce the field strength below the pair production limit (\ref{Ec}).
Assuming $\mathbf{B}=0$, the energy of the field is given by \cite{Tensor1, Tensor2} 
\begin{equation}
{\cal E}[E]=\int\left\{ \frac{E^{2}}{8\pi}+4\eta E^{4}+...\right\} \, d^{3}x<2m_{e},\label{weak3}
\end{equation}
where the first term in (\ref{weak3}) is the energy of the field
as given by classical electrodynamics and the second term is a correction
due to the non-linearity of the new Maxwell's equations given by the Euler-Heisenberg 
term (\ref{L3}).
A sufficient condition for the absence of pair creation is then given by, 
\begin{equation}
{\cal E}[E]<2m_{e}.
\end{equation}

We will come back to these conditions after solving for particular
electric fields. 

A valid question is whether we should keep our results at the most
quadratic in $e$ since our Lagrangian is up to this order. Answering
this question we should not forget that the theory is a self-interacting
one and bears a strong formal similarity to electrodynamics in matter.
In analogy to the Euler-Heisenberg theory let us consider within
Maxwell theory (whose Lagrangian is linear in $e$) the electromagnetic
waves in conducting matter. Ohm's law $\mathbf{j}=\sigma\mathbf{E}$
together with the Maxwell equations leads to the telegraph equation
whose solution is $\mathbf{E}=\mathbf{E}_{0}e^{(i\mathbf{k}\cdot\mathbf{r}-\omega t)}$.
The dispersion relation reads $k=|\mathbf{k}|=k(\sigma(\omega),\omega)$
in which one includes the frequency dependence of the conductivity
$\sigma=\sigma_{0}/(1-i\omega\tau)$. The damping time $\tau$ depends
on the charge via, $\tau=m\sigma_{0}/ne^{2}$. The conductivity $\omega$
enters $k$ in a non-linear way which in turn enters the solution.
We do not expand neither the conductivity nor the solution in the
charge albeit the Lagrangian (and the damping equations which model
the damping) are linear in $e$. If it is not stringent to expand
our results in $e$, it means that we can retain the full non-linearity
of the theory \cite{greiner}. 
Doing it we should re-consider the meaning of perturbation
for the following reason. In the static case of the Euler-Heisenberg
theory the deviations from the Maxwellian field $\mathbf{E}_{0}$
at small distances can be large. Certainly, these large deviations
are not corrections to $\mathbf{E}_{0}$. But this is the case also
for electric potential of the proton where at short distances the
proton potential is not even approximately Coulomb and nonetheless
this finite size correction is considered a perturbation as far as
physical observables, such as the energy level, are concerned. Basically,
the nature of perturbation in such a case is due to the very short range
of the new effect. In the Euler-Heisenberg theory happens exactly
the same: the large deviations are restricted to a very small range
and therefore appear as perturbations in the observables.
By keeping the full dependence in $r$ at the short range we
follow here the same procedure used also in \cite{Tensor1,Tensor2}.

\subsection{The electrostatic case}

In contrast to the Maxwell case, where in the static case the equations
for the electric and magnetic field decouple, in the Euler-Heisenberg
theory we have to set $\mathbf{B}=0$ by hand if we want a pure electrostatic
case. So, let us now set $\mathbf{B}=0$ and neglect time derivatives
in the new Maxwell's equations.

The Gauss law for $\mathbf{D}$ allows us to make a direct connection
to the Maxwellian field $\mathbf{E}_{0}$ via 
\begin{equation}
{\mathbf{\nabla}}\cdot{\mathbf{D}}=\rho={\mathbf{\nabla}}\cdot{\mathbf{E}}_{0}\label{stat1}
\end{equation}
Hence we end up with an algebraic equation of the form 
\begin{equation}
{\mathbf{E}}+4\pi{\mathbf{P}}={\mathbf{E}}+\tilde{\eta}_{1}E^{2}{\mathbf{E}}={\mathbf{E}}_{0}\label{stat2}
\end{equation}
where 
\begin{equation}
\tilde{\eta}_{1}=16\pi\eta=\frac{2\alpha^{2}}{45\pi m_{e}^{4}}.\label{stat3}
\end{equation}
For the spherically symmetric case we obtain a third order polynomial
equation for the magnitude of the electric field $E$ given the ``standard''
Maxwellian field $E_{0}$: 
\begin{equation}
E+\tilde{\eta}_{1}E^{3}=E_{0}\label{stat4}
\end{equation}
The solution of the third order algebraic equation (\ref{stat3})
is always dependent on $E_{0}$, i.e., $E[E_{0}]$. The only real
solution to (\ref{stat4}) is given by Cardano's formula 
\begin{equation}
E=\left(\frac{E_{0}}{2\tilde{\eta}_{1}}\right)^{1/3}\times\left(\sqrt[3]{1+\sqrt{1+\frac{4}{27\tilde{\eta}_{1}E_{0}^{2}}}}
+\sqrt[3]{1-\sqrt{1+\frac{4}{27\tilde{\eta}_{1}E_{0}^{2}}}}\right).\label{stat5}
\end{equation}
As advertised above, for strong fields $E_{0}$, the Euler-Heisenberg field E given by (\ref{stat4})  behaves as $E\propto(E_{0})^{1/3}$ (this amounts to retaining 
the cubic term in the polynomial equation (\ref{stat4})), i.e., the field becomes
weaker as compared to the classical Maxwellian counterpart. The above actually is true no matter the value of $E_{0}$, but the reduction of the field strength is more pronounced the bigger the value of $E_{0}$ is.
Do note that, in accordance with the last paragraph of the previous section, the result (\ref{stat5}) is not a perturbative series in the fine structure constant.

The case of a point charge, where $E_{0}=\frac{\sqrt{\alpha}}{r^2}$, is 
worked out in References \cite{Tensor1, Tensor2}. It is found there that the energy of a point charge is ${\cal E}[E]=2.09m_e$ after the $\gamma-\gamma$ corrections
are included: a finite result that contrasts with the classical electrodynamics case. 
Do note that the energy of the point charge barely surpasses two times the electron mass. 
We will see that for our model for the proton the resulting electrostatic energy 
is below the pair production threshold.

A valid question is what will happen if we include terms of higher
orders in $\alpha$ in the Lagrangian. To this end we can do two things.
First, we can choose to keep more terms in the weak field expansion
of (\ref{EH83}) (see appendix). Secondly, we can take the contribution to the effective
Lagrangian due to higher loops, specifically, 
the two loops Euler-Heisenberg Lagrangian \cite{kors}. 
The contribution of the second loop to the Euler-Heisenberg Lagrangian reads 
\cite{weakfieldL,kors}, 
\begin{eqnarray}
{\cal L}_{EH}^{(2)}&=&\frac{\alpha^{3}}{64\pi^{4}m_{e}^{4}}\left(\frac{16(\mathbf{E}^{2}-\mathbf{B}^{2})^{2}}{81}+\frac{263(\mathbf{E}\cdot\mathbf{B})^{2}}{162}\right)
\nonumber \\
&-&\frac{\alpha^{4}}{256\pi^{4}m_{e}^{8}}\left(\frac{1219(\mathbf{E}^{2}-\mathbf{B}^{2})^{3}}{2025}+\frac{8656(\mathbf{E}^{2}-\mathbf{B}^{2})(\mathbf{E}\cdot\mathbf{B})^{2})}{2025}\right).\label{2loop}
\end{eqnarray}
Specializing, as before, to the case of $\mathbf{B}=0$,the  Lagrangian
(\ref{2loop}) reduces to 
\begin{equation}
{\cal L}_{EH}^{(1+2)}={\cal L}_{EH}^{(1)}+{\cal L}_{EH}^{(2)}=\eta_{1}\mathbf{E}^{4}+\eta_{2}\mathbf{E}^{6}\label{2loops2}
\end{equation}
with 
\begin{eqnarray}\label{etacoeffs} 
\eta_{1} & = & \frac{\alpha^{2}}{360\pi^{2}m_{e}^{4}}+\frac{\alpha^{3}}{324\pi^{4}m_{e}^{4}}\nonumber \\
\eta_{2} & = & \frac{\alpha^{3}}{630\pi^{2}m_{e}^{8}}-\frac{1219\alpha^{4}}{518400
\pi^{4}m_{e}^{8}}
\end{eqnarray}
The algebraic version of the Gauss's law is now a quintic polynomial
\begin{equation}
E+\tilde{\eta}_{1}E^{3}+\tilde{\eta}_{2}E^{5}=E_{0},\label{quintic}
\end{equation}
where
\begin{eqnarray}
\tilde{\eta}_{1} & = & 16\pi\eta_{1},\nonumber \\
\tilde{\eta}_{2} & = & 24\pi\eta_{2}.
\end{eqnarray}
We shall refer to the expansion of the polynomial equation as order (e.g. the third 
order polynomial, $E+{\eta}_{1}E^{3} = E_0$ is called first order since it is the first 
correction of the Euler-Heisenberg theory). We also distinguish the loop contribution 
and for instance, Eq.(\ref{quintic}) would be second order with two loops.
Let us note that the contribution from the two loops Lagrangian amounts to a small correction to the one loop coefficients, hence it can be safely ignored in the first approximation. In any case, for strong classical fields  $E_{0}$, the Euler-Heisenberg field E given by (\ref{quintic})  behaves as $E\propto(E_{0})^{1/5}$. The result is that by taking more and more terms in the weak field expansion of the Euler-Heisenberg Lagrangian, the resulting field strength gets smaller and smaller compared to the classical field given by the Maxwell's equations.
We can continue taking more and more terms from the weak field expansion. For example, taking four terms the equation for the electric fields becomes
\begin{equation}
E+\tilde{\eta}_{1}E^{3}+\tilde{\eta}_{2}E^{5}+\tilde{\eta}_{3}E^{7}+\tilde{\eta}_{4}E^{9}+\ldots=\sum_{i=0}\tilde{\eta}_i  E^{2i+1}=E_{0}\label{9 order}
\end{equation}
where $\tilde{\eta}_0=1$ and, taking into account only the one-loop Lagrangian, the third and fourth coefficients
are (see appendix)
\begin{align}
\tilde{\eta}_{3} & =\frac{32\alpha^{4}}{315\pi m_{e}^{12}},\nonumber \\
\tilde{\eta}_{4} & =\frac{160\alpha^{5}}{297\pi m_{e}^{16}}.
\end{align}
In the case of the proton, the maximal value of $E_0$ is $5 \times 10^3\,  MeV$.
In this case the field will behave as $E\propto(E_{0})^{1/9}$, and in general, taking $n$ terms in the weak field expansion of the Lagrangian results in a field behaving as $E\propto(E_{0})^{1/2n+1}$. However, as the weak field expansion for the Euler-Heisenberg Lagrangian is an asymptotic series, it does not make much sense to try to take
much more terms in (\ref{9 order}).
We will address this point in more detail in a subsequent section.
However, already here we can exemplify the effect of taking more terms in the 
Lagrangian expansion for a Coulomb-like field. Eq. (\ref{stat5}) applied to 
the classical Coulomb field $E_0 \simeq r^{-2}$ gives at short distances a field 
that behaves like $E \simeq r^{-2/3}$, which, together with the condition 
(\ref{conditions}) results in the range of validity of $r >$ 257 fm. If we take into 
account the next correction to the Gauss law given by Eq. (\ref{quintic}), the 
resulting electric field at short distances behaves like $E \simeq r^{-2/5}$ which then 
leads to $r >$ 154 fm. This shows how taking into account the corrections from the 
Euler-Heisenberg expansion results in electric fields that behave better than the 
classical ones.  

\section{Results}

Below we will present the results in the form of figures since most of
the equations cannot be handled analytically. Let us start by showing
the effect of the $\gamma\gamma$ interaction (Euler-Heisenberg) on
the electric field of the proton. In Fig. 6 we have plotted $E_{p}$
with the finite size corrections (i.e., with the inclusion of the
form-factors (FF)) in the Maxwell theory and in the Euler-Heisenberg
theory (two loops). The Darwin term refers here to the proton
Darwin term. The effect is clear: a reduction of the field strength
at small distances.

\begin{figure}[h]
\includegraphics[width=8cm,height=8cm]{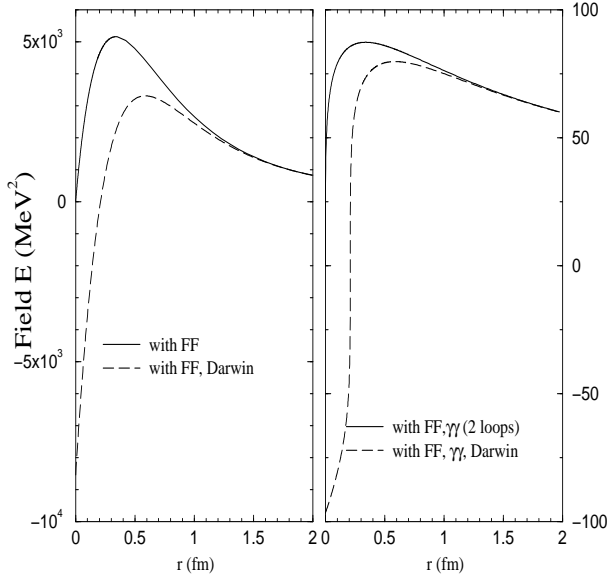}
\caption{Effect of the Euler-Heisenberg theory (right panel) on the electric field
of the proton (with form factors included). 
By two loops we mean, second order 
(5$^{th}$ order polynomial as in Eq. (\ref{quintic}))    
with two loop corrections.} 
\end{figure}

\begin{figure}[h]
\includegraphics[width=8cm,height=8cm]{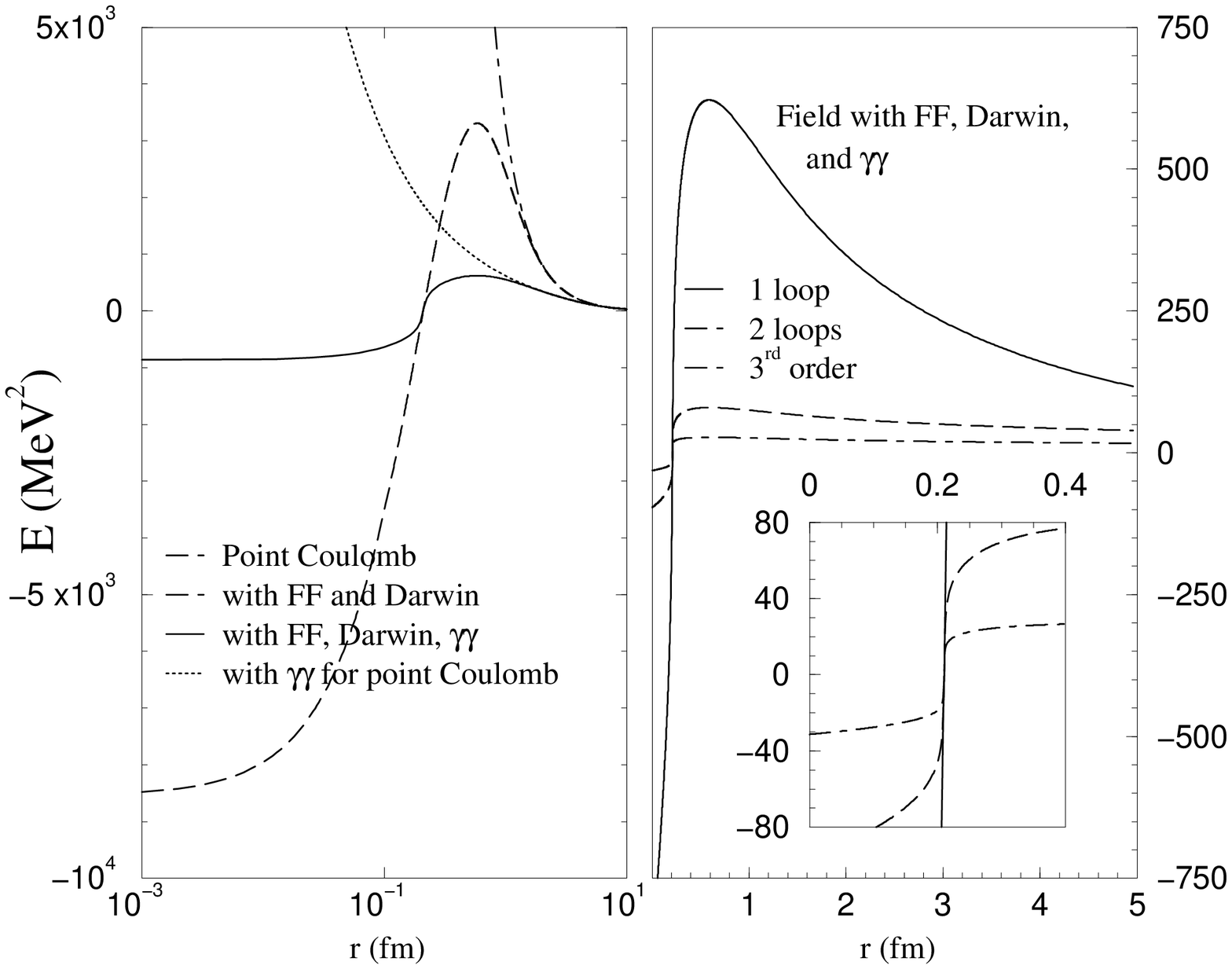}
\caption{Comparison of the effect of the Euler-Heisenberg theory on
the electric field of the proton. We compare with the point-like
proton (left panel) and demonstrate the effects of higher orders (right panel). 
The Darwin term
refers to proton Darwin. Again, by two loop we mean 2$^{nd}$ order with two
loops (Eq.(\ref{quintic})). This will be the same in the subsequent figures.}
\end{figure}

In Fig. 7 (left panel) we show a comparison between the electric field of the 
point-like proton and the
extended proton after the $\gamma\gamma$ corrections at one loop are included.
We also compare the effect of higher orders by the inclusion of the
hypothetical (see text) third order result (right panel). The saturation of the result
is visible: the reduction of the field strength is less with higher
order. The inlay demonstrates the short range behavior. 
It is also obvious that the fluctuations at distances around 1 fm become much 
less prominent with higher orders of the expansion included. The inclusion of 
every higher order brings us closer to satisfying the condition in (\ref{conditions}).

\begin{figure}[h]
\includegraphics[width=8cm,height=8cm]{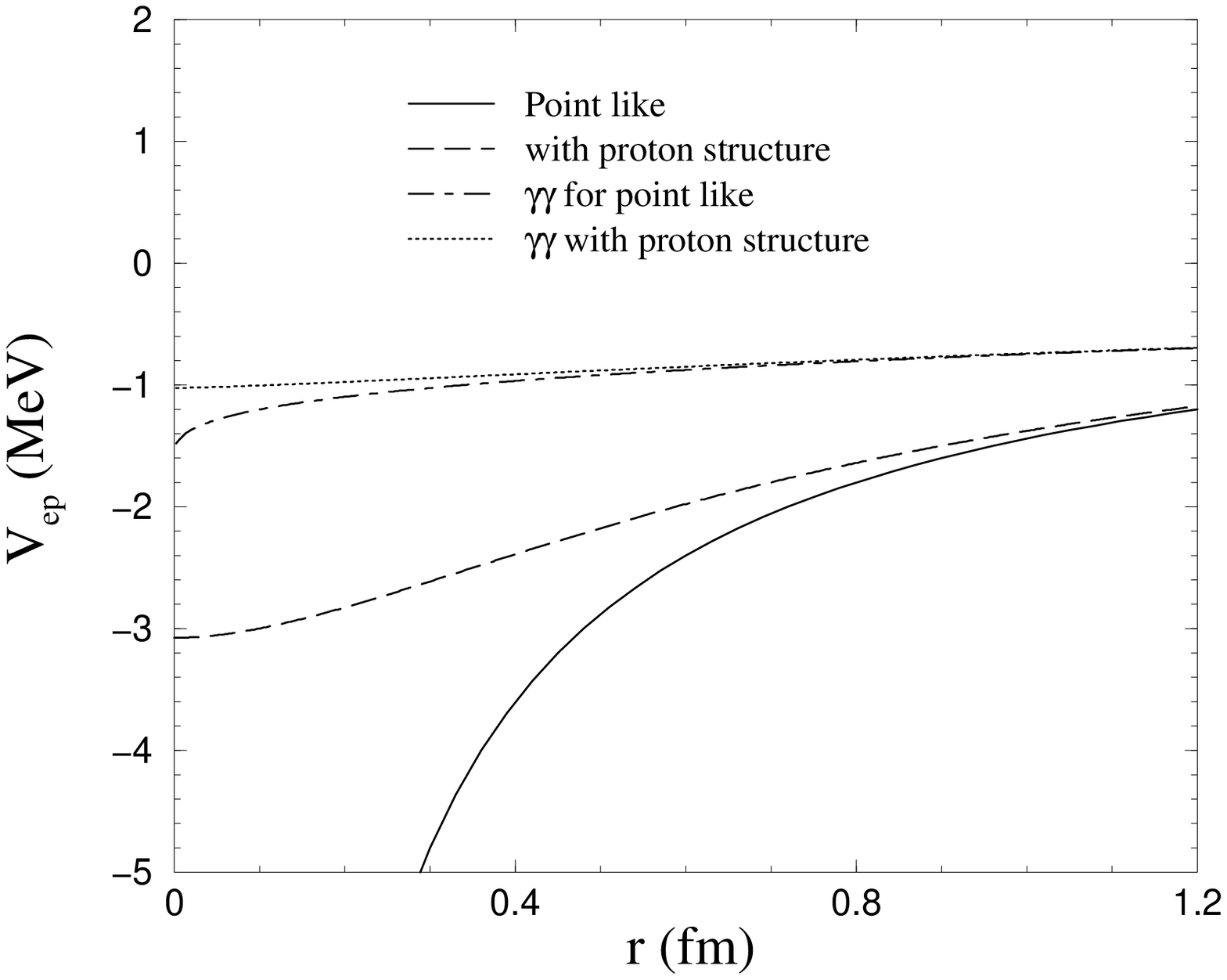}
\caption{The electron proton interaction potential with and without the $\gamma\gamma$
correction at one loop. The Darwin terms are not included here.}
\end{figure}

We can do a similar comparison for the electron-proton interaction
potential $eV_{p}$ on the one hand and $eV_{p}^{\gamma\gamma}$ on
the other hand. This calculation (at one loop) 
can be found in Fig. 8 without the inclusion
of the electron or proton Darwin term. To understand the effect of the latter,
let us recall that the end result will matter depending on whether we
handle the electron Darwin term within the Dirac theory or the Breit
formalism. In the Dirac theory the full interaction potential after
the Euler-Heisenberg corrections would be given by 
\begin{equation}
V_{ep}^{\gamma\gamma}({\rm Dirac})=-e(V_{C}+V_{pD})^{\gamma\gamma}+\frac{e}{8m_{e}^{2}}\nabla\cdot\mathbf{E}^{\gamma\gamma}\label{Dirac1}
\end{equation}
This is to be contrasted with the result within the Breit formalism
which gives 
\begin{equation}
V_{ep}^{\gamma\gamma}({\rm Breit})=-e(V_{C}+V_{pD})^{\gamma\gamma}+
\frac{e}{8m_{e}^{2}}\rho_{C}\label{Breit1}
\end{equation}
We recall that in the Breit convention we use $\rho_{C}=e\rho$. 
Fig. 9 summarizes the effect of the $\gamma\gamma$ correction within the
Dirac formalism as given in Eq. (\ref{Dirac1}), where, $V_{pD}$ corresponds to 
the proton Darwin term and $\frac{e}{8m_{e}^{2}}\nabla\cdot\mathbf{E}^{\gamma\gamma}$ 
corresponds to the electron Darwin term. In the Breit formalism (Eq. (\ref{Breit1})), 
$\frac{e}{8m_{e}^{2}}\rho_{C}$ corresponds to the electron Darwin term.

\begin{figure}[h]
\includegraphics[width=8cm,height=8cm]{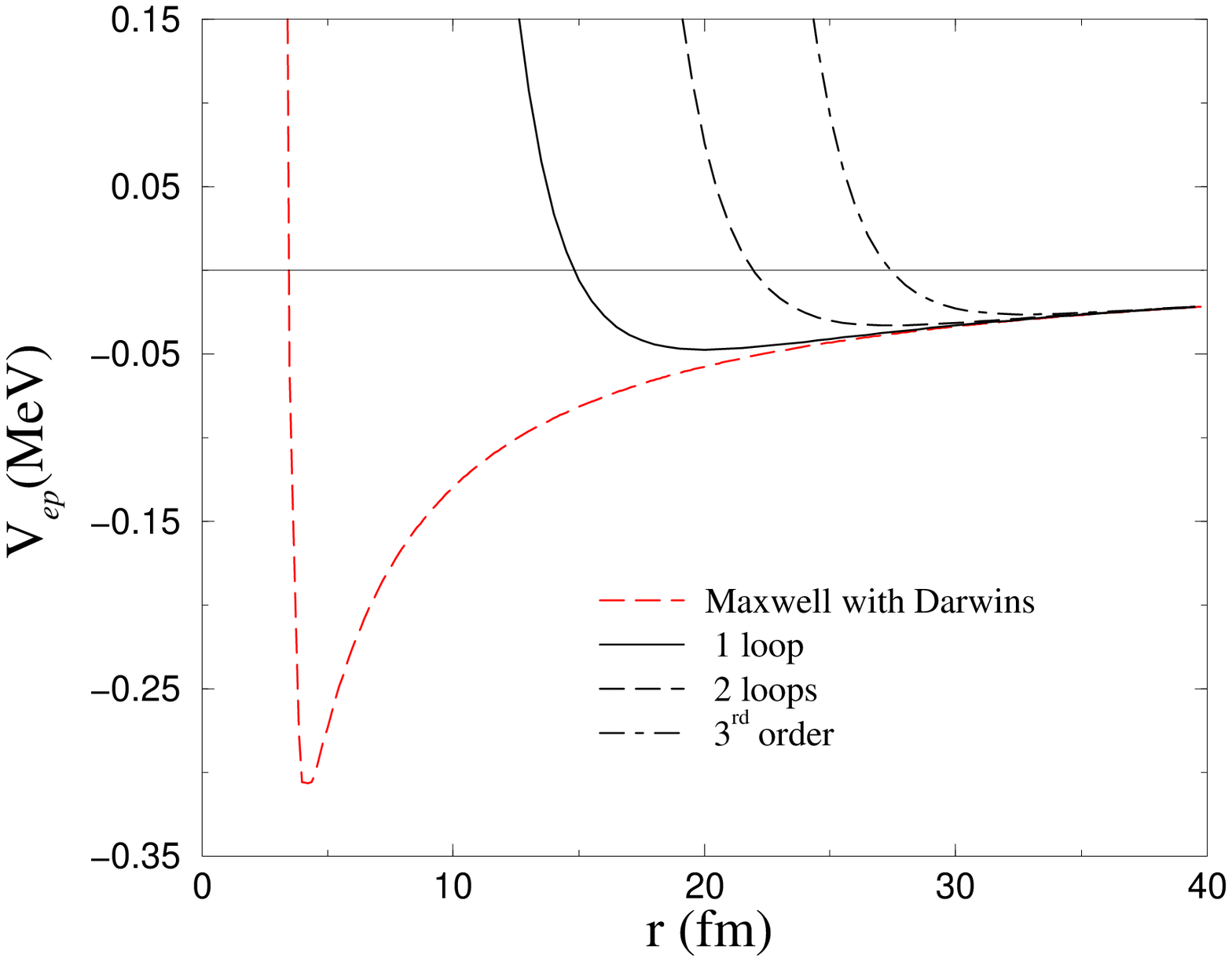}
\caption{Effect of the Euler-Heisenberg theory on the electron proton 
interaction potential within the Dirac theory (see Eq. (\ref{Dirac1}) and 
text).}
\end{figure}

Fig. 10 does the same for the Breit Hamiltonian (\ref{Breit1}) where
in addition we display the effects of the electron Darwin term proportional
to $\rho$. Note that the effects of the proton Darwin become visible only in 
a very small range of $r$ as will be seen in Fig. 12.
Finally, Fig. 11 compares the Dirac (\ref{Dirac1}) and
the Breit Hamiltonian (\ref{Breit1}).

\begin{figure}[h]
\includegraphics[width=8cm,height=7cm]{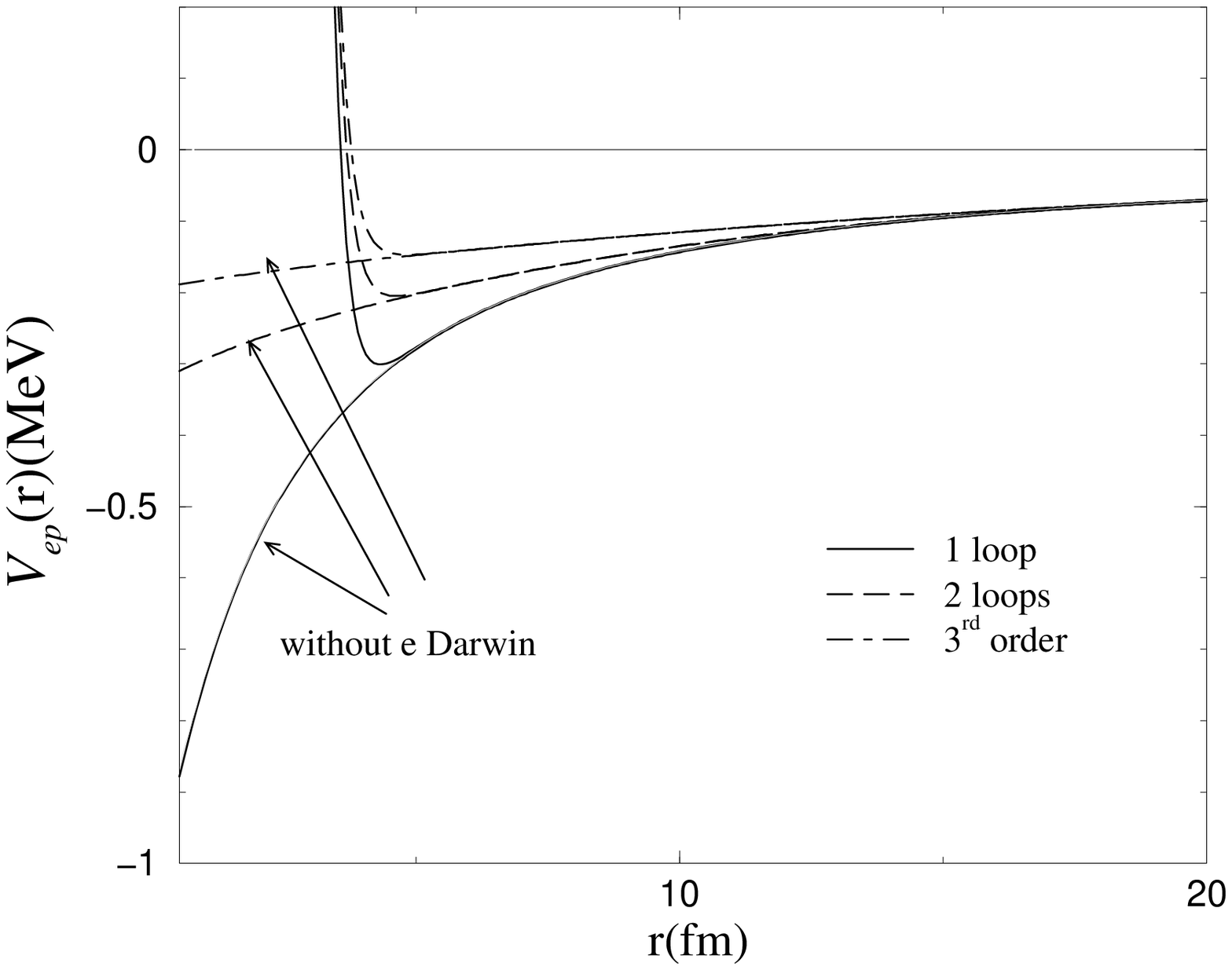}
\caption{Effect of the Euler-Heisenberg theory on the electron proton interaction
potential within the Breit theory (see Eq. (\ref{Breit1}) and text).}
\end{figure}

\begin{figure}[h]
\includegraphics[width=8cm,height=8cm]{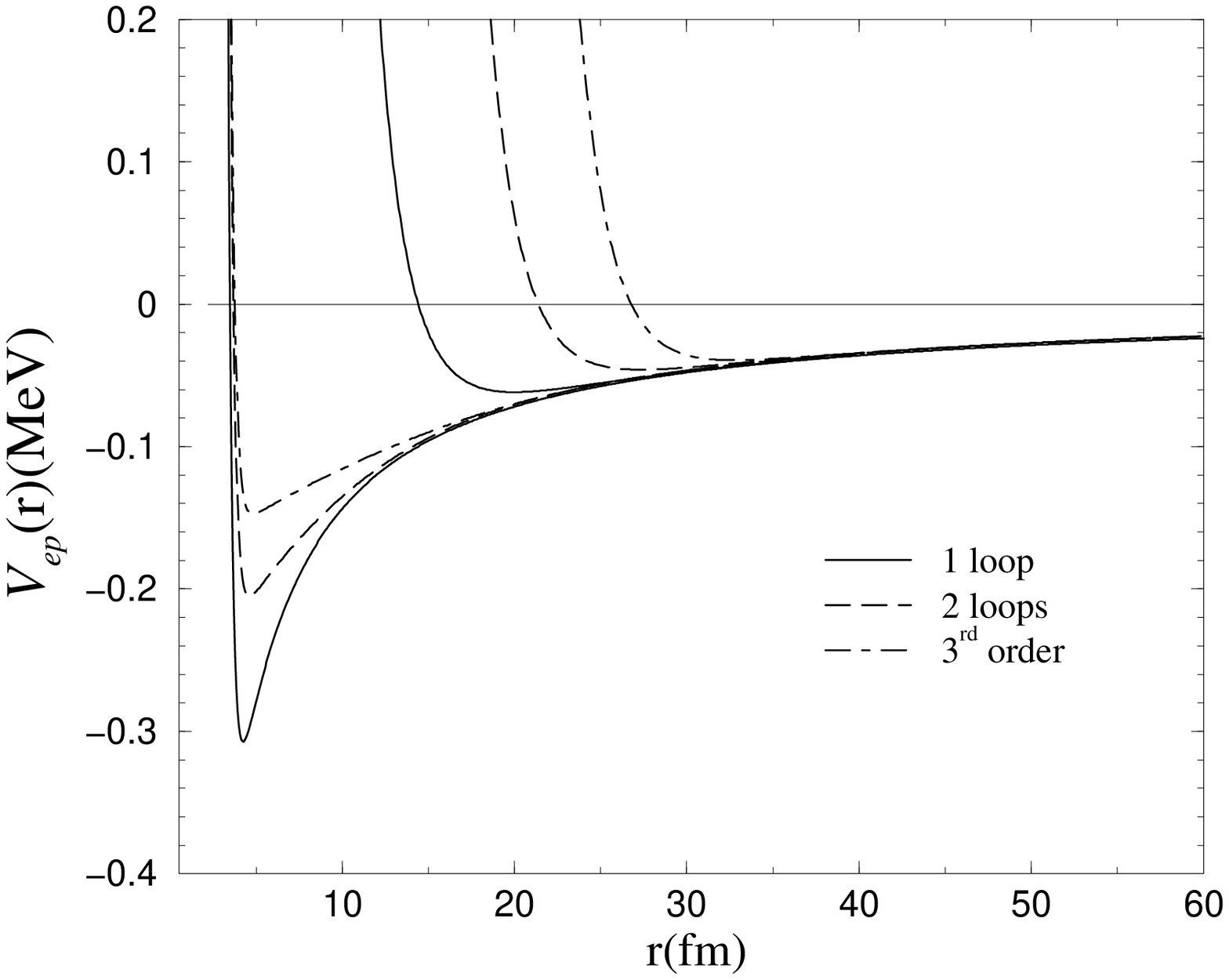}
\caption{Comparison of the effects of the Euler-Heisenberg theory on
the electron proton interaction potential within the Dirac and Breit formalisms
(see text). From the left to right, the first three curves correspond
to the Breit Hamiltonian and the next three to the Dirac theory.}
\end{figure}

Coming back to the Dirac formalism we can also examine the effect
of the proton Darwin term at small distances. In spite of being suppressed
by the proton mass squared this term has an effect at small distances.
Fig. 12 summarizes our findings here. Fig. 13 does the same for the
Breit formalism. In both figures the electron Darwin term is included.

\begin{figure}[h]
\includegraphics[width=8cm,height=8cm]{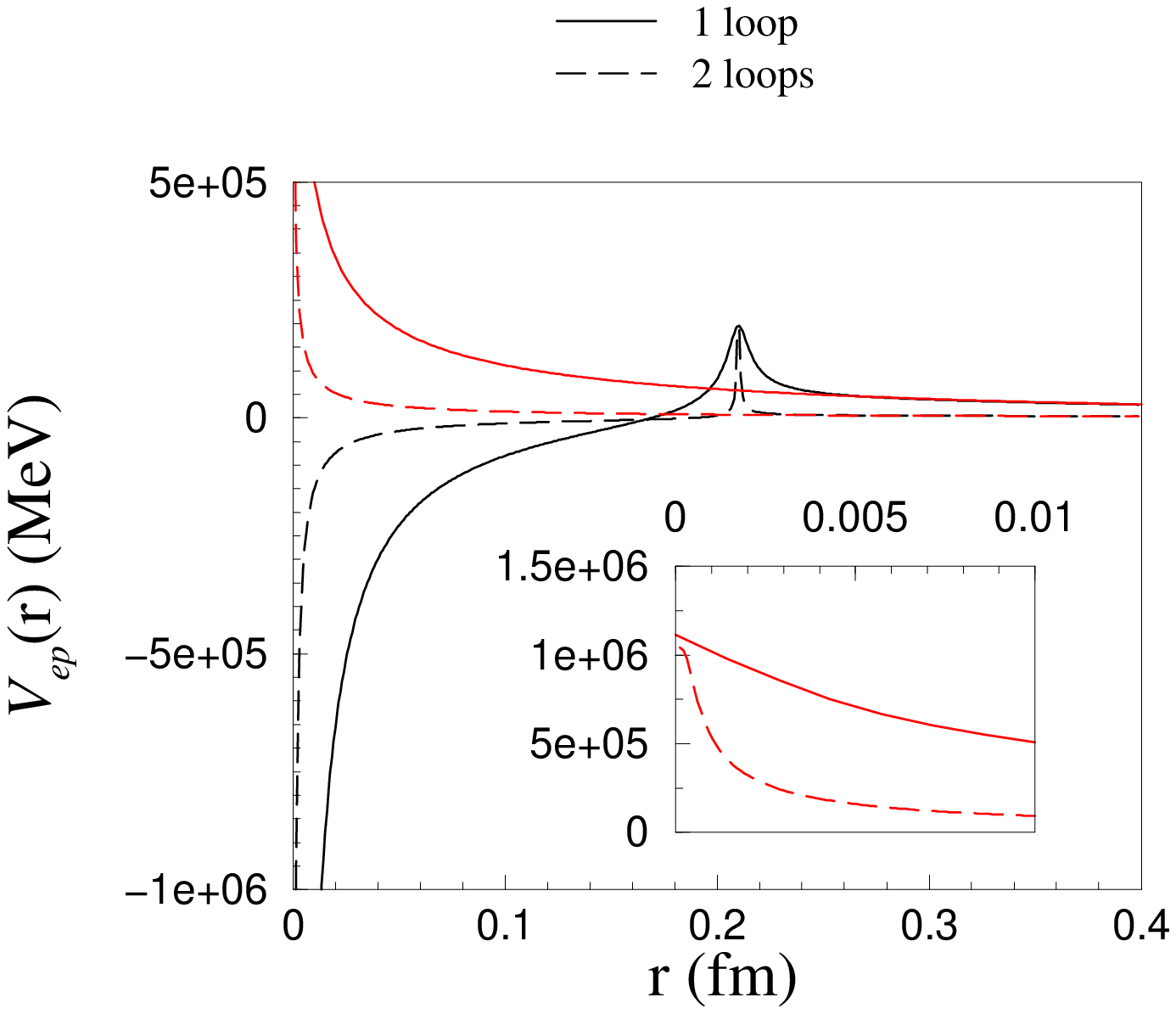}
\caption{Effect of the Euler-Heisenberg theory on the electron proton 
interaction potential within the Dirac theory (see text). We focus
here on the effect of the proton Darwin term at small distances. The red
curves are without the proton Darwin term whereas the black ones include
this contribution.}
\end{figure}

\begin{figure}[h]
\includegraphics[width=8cm,height=8cm]{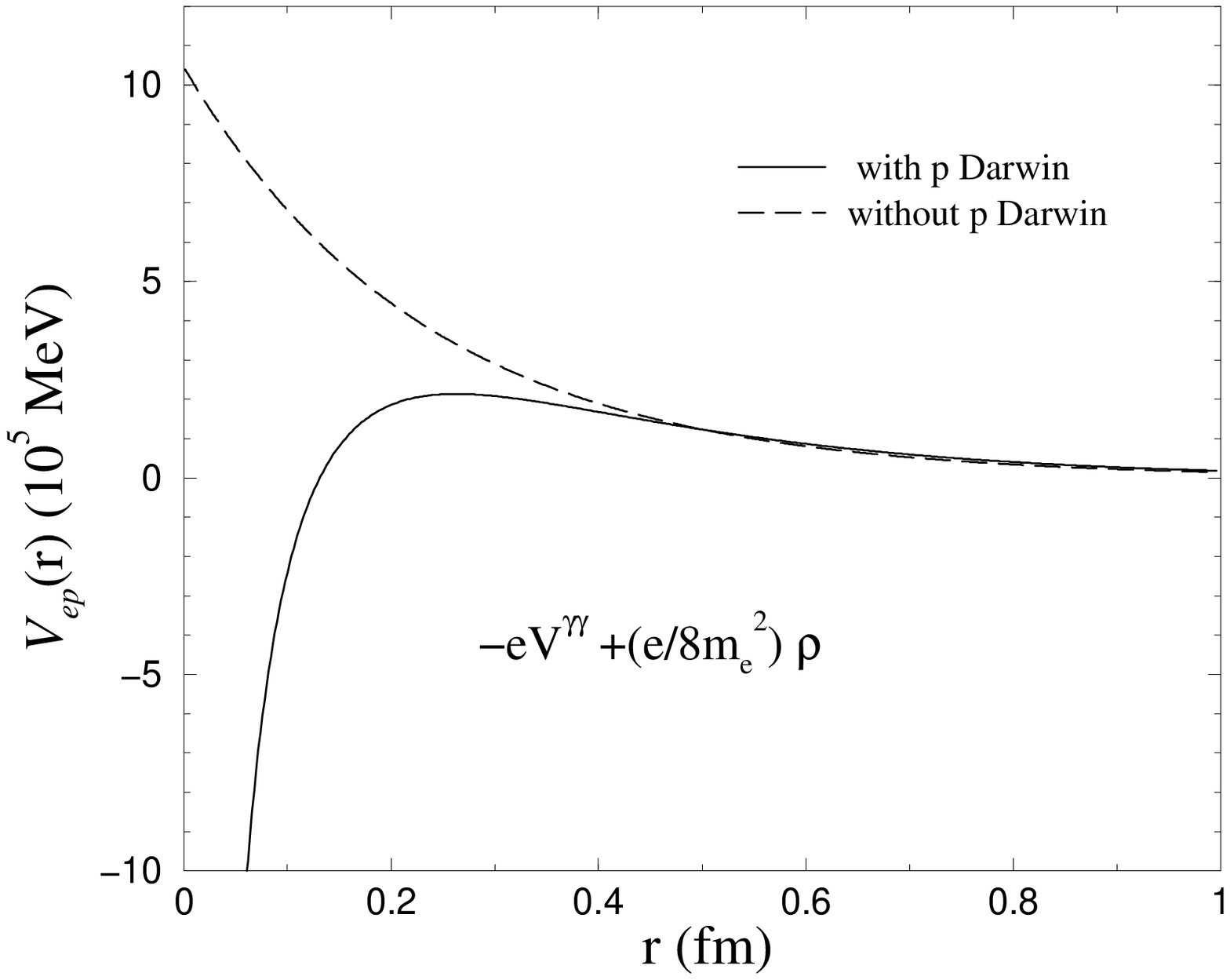}
\caption{Effect of the proton Darwin term on the electron proton
interaction potential within the Breit theory (see text) using the Euler-Heisenberg 
at one loop.}
\end{figure}

The difference between the the potential with and without electron
Darwin term in the Dirac theory is displayed in Fig. 14. In order to see 
this difference, we have not included the proton Darwin terms in this plot. 

\begin{figure}[h]
\includegraphics[width=8cm,height=10cm]{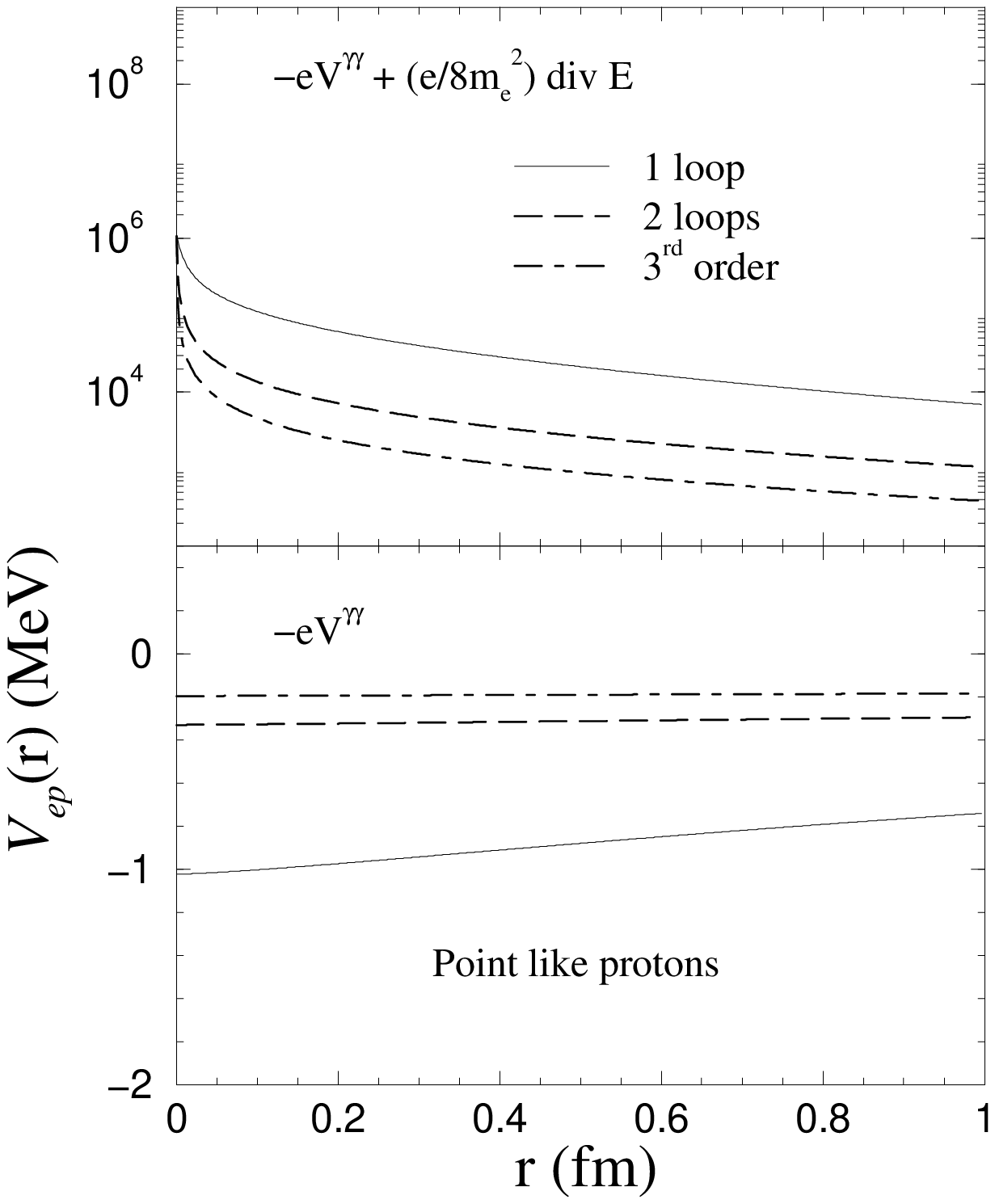}
\caption{Effect of the electron Darwin term on the electron proton interaction 
potential within the Euler Heisenberg theory with the Dirac formalism (see text). 
Note that the proton Darwin is not included. 
}
\end{figure}

Finally, we recall that the energy content of the electric field is
${\cal E}[E]=(1/8 \pi)\int E^{2}d^{3}x$. Applying it to the proton
field $E_{p}$ without the Euler-Heisenberg corrections we obtained
${\cal E}[E_{p}]\simeq\,\,1 MeV$. After the corrections this number
reduces considerably 
\begin{eqnarray}
{\cal E}[E_{\gamma\gamma,1-loop}] & \simeq & 0.46\,\, MeV\nonumber \\
{\cal E}[E_{\gamma\gamma,2-loop}] & \simeq & 0.19\,\, MeV
\end{eqnarray}
This includes corrections to the energy-momentum tensor (see appendix 
for details).
One can see that the loop results are well below the pair production
threshold field.

\section{Static electric fields in the Euler-Heisenberg theory}
 The full Euler-Heisenberg Lagrangian, in principle, contains information also
on the strong fields. For static electric fields generated by a single source, 
the weak field expansion of
this Lagrangian gives a polynomial equation for the electric field $E$ whose order
$N$ indicates the truncation of the expansion. In general, we have
\begin{equation} \label{me1}
E +\sum_{i=1}^N \tilde{\eta}_{i} E^{2i+1}=E_0
\end{equation}
where $E_0$ is the result of computing the electric field in the Maxwell theory
assuming static charge distribution. The coefficients $\tilde{\eta}_i$ are listed in the
appendix where one can see that they have the structure $\tilde{\eta}_i=n_i\frac{\alpha^{i+1}}{\pi m_e^{4i}}$
with $n_i$ being purely numerical factors. Effects of higher loops should, in principle, enter the coefficients
$\tilde{\eta}_i$, but this will only correct these coefficients as was evident from the two-loops example.
We have already shown that the effect of the polynomial equation is to reduce the value $E_0$ when starting with
a large electric field. This implies that the Euler-Heisenberg reduces the electric field strength automatically below
the pair production threshold. In this context arises the obvious question, 
up to which order $N$ should one continue the
expansion.  Since we deal here with non-convergent series a reasonable starting assumption is to insist on a reduction such that equation (\ref{Ec})
is satisfied.  To probe more into this matter we re-write equation (\ref{me1}) as
\begin{equation} \label{me2}
\xi +\sum_{i=1}^N \tilde{n}_i\xi^{2i+1} =\xi_0 \equiv \frac{E_0\sqrt{\alpha}}{m_e^2}
\end{equation}
with the dimensionless variable
\begin{equation} \label{me3}
\xi \equiv \frac{E\sqrt{\alpha}}{m_e^2}
\end{equation}
and $\tilde{n}_i=(n_i \alpha)/\pi$. 
In general, $\xi$ will, of course depend on $r$ if $\xi_0$ is space dependent. What we have in mind is to choose a constant
$\xi_0$ corresponding to the maximal field strength. In the case of the proton this value is $5\times 10^3\, MeV$ which implies
that we have $\xi_0=1700$. We look numerically for $\bar{N}$ such that $\xi < 1$. The result can be found in Fig. 15. 
\begin{figure}[h]
\includegraphics[width=7cm,height=7cm]{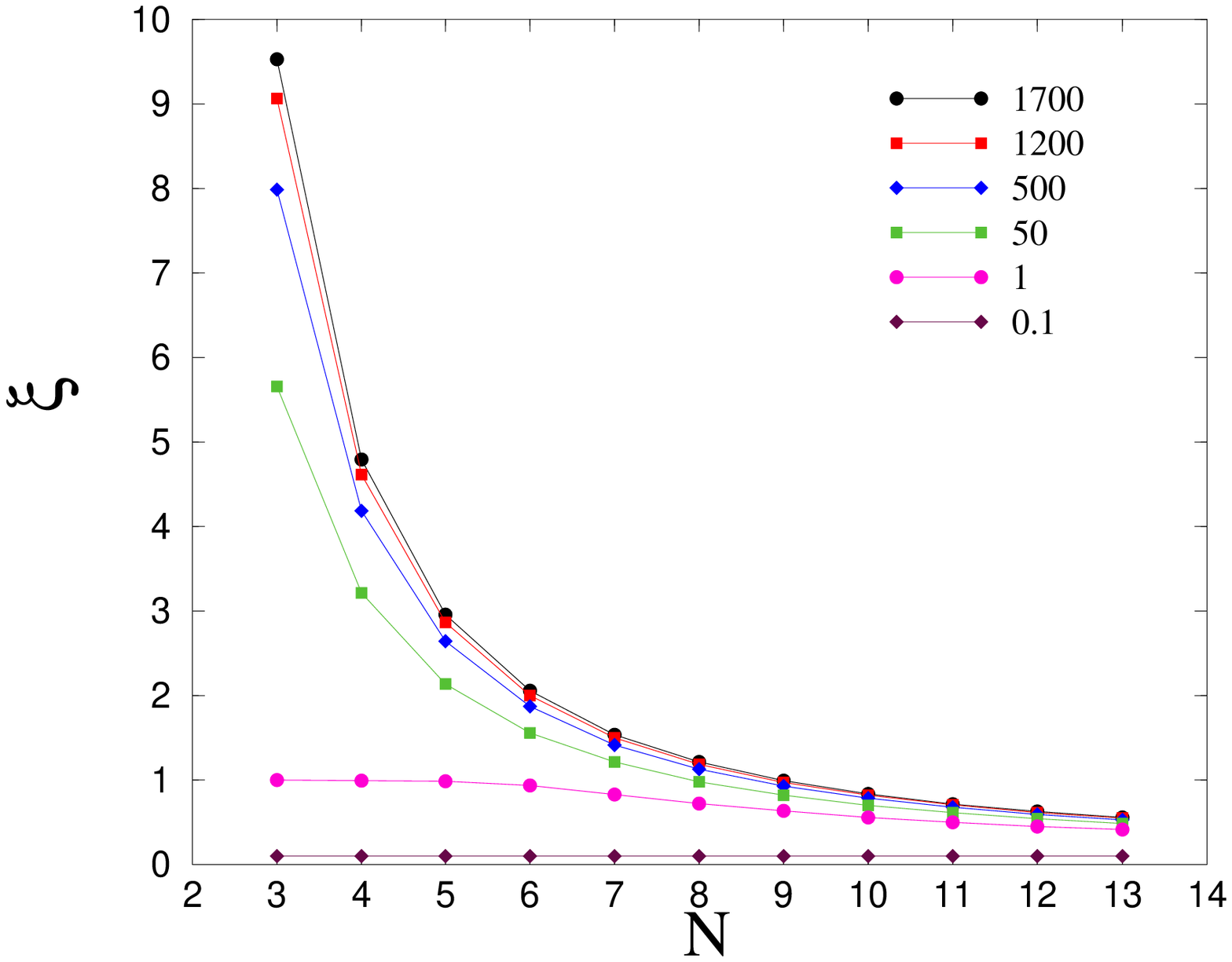}
\caption{Effects of higher orders on the dimensionless variable 
$\xi \equiv E\sqrt{\alpha}/{m_e^2}$.
}
\end{figure}
For $\bar{N}=9$ we obtain $\xi=0.99$. Choosing another $\xi_0 > 1$ gives globally the same result. It is only if we start
with $\xi_0 <1$ can we treat equation (\ref{me1}) fully perturbatively. For strong fields we should expand up to order $\bar{N}$
such that the next orders starting with $\bar{N}+1$ are small and can be treated perturbatively. Indeed, for an observable
like the energy-momentum tensor we can write
\begin{equation} \label{me4}
T_{00}=\sum_i \frac{k_i}{\pi^2}m_e^4 \xi^{2i}
\end{equation}
where $k_i$ are purely numerical coefficients of the order $1$ (see appendix). In such a case the above sum,
albeit not convergent, makes sense if we have $\xi < 1$.

\section{Conclusions}
It is not often that we encounter in physics extreme strong electric fields, fields such that
$\xi=E\sqrt{\alpha}/m_e^2 \gg 1$ and with an energy content which would, in principle, allow pair production. Using
Maxwell's electrodynamics, such an electric field felt by the electron in the hydrogen atom seems possible due
to the finite size effects of the proton. Not only are we at the threshold of pair production (the exact number
of the energy content will depend on the parametrization of the from factors), but other 
strange phenomena like a 
positron-proton bound state (or, in general, a positive energy Gamow state) seem possible. A valid question arises
whether such strong fields are indeed present at the short distances in the 
hydrogen atom. Strong electromagnetic fields have been of interest for quite 
some time (see e.g. \cite{jan1,jan2,jan3}) and can be applied in different 
contexts \cite{jan4}.  
We have demonstrated that going
from Maxwell's electrodynamics to the Euler-Heisenberg theory, the strong electric 
field will be reduced automatically far below
the pair production threshold, at least as far as the energy content of the 
field is concerned. At the same time, the potential as seen by a positron 
will be too flat to allow bound states. To reach this effect we have to
expand the original Euler-Heisenberg Lagrangian in higher order in $\alpha$. 
We can consistently use some terms in the expansion 
by the requirement $\xi < 1$ even though the expansion is only asymptotic. 
We emphasize that fields resulting from the Euler-Heisenberg corrections 
cannot be viewed as small corrections to fields calculated with the Maxwell theory. 
However, due to the short range in which the Euler-Heisenberg results differ 
significantly from the Maxwell's ones, it is possible to treat the change in the 
observables perturbatively. 
Similarly, before we reach $\xi < 1$, every order
in the expansion in the Euler-Heisenberg Lagrangian is not a correction to the 
original Maxwellian field (as far as its strength is concerned), however, 
this happens at a very short range.
However, whenever $\xi < 1$, i.e., from an order on at which this condition is satisfied we can talk about perturbations in the expansion of the
observables like the energy-momentum tensor.     

\section*{Author Contribution Statement}
All authors have contributed equally to the discussions, calculations and 
preparation of the manuscript. 
\section*{Appendix}

In a purely electric field the Euler-Heisenberg Lagrangian reduces
to

\begin{equation}
\mathcal{L}_{EH}=-\frac{1}{8\pi^{2}}\int_{0}^{\infty} \,ds \,
\frac{e^{-sm_{e}^{2}}}{s^{3}}\left[(eEs)\cot(eEs)-1+\frac{1}{3} e^2 
E^{2}s^{2}\right].\label{A1}
\end{equation}

With the help of the Taylor expansion

\begin{eqnarray}
(eEs)\cot(eEs) & = & 1-\frac{1}{3}(e^{2}s^{2})E^{2}-\frac{1}{45}(e^{4}s^{4})E^{4}-\frac{2}{945}(e^{6}s^{6})E^{6}\nonumber \\
 &  & -\frac{1}{4725}e^{8}s^{8}a^{8}-\frac{2}{93555}e^{10}s^{10}E^{10}+\ldots,\label{A2}
\end{eqnarray}
the Lagrangian (\ref{A1}) can be expanded as

\begin{equation}
\mathcal{L}_{EH}=-\frac{m_{e}^{4}}{8\pi^{2}}\sum_{n=0}^{\infty}\frac{(-1)^{n}\mathcal{B}_{2n+4}}{(2n+4)(2n+3)(2n+2)}\left(\frac{2eE}{m_{e}^{2}}\right)^{2n+4}\label{A3}
=\sum_{n=0} a_n E^{2n+4}
\end{equation}
where $\mathcal{B}_{2n+4}$ are Bernoulli numbers.

The series (\ref{A3}) is divergent, non-alternating and is not Borel
summable. This divergence is related to the possibility of pair production
due to an electric field. However, for low field strengths, we can
take the first terms of the expansion as an approximation. The first
four terms of (\ref{A3}) are

\begin{equation}
\mathcal{L}_{EH}=\frac{\alpha^{2}}{360\pi^{2}m_{e}^{4}}E^{4}+\frac{\alpha^{3}}{630\pi^{2}m_{e}^{8}}E^{6}+\frac{\alpha^{4}}{315\pi^{2}m_{e}^{12}}E^{8}+\frac{4\alpha^{5}}{297\pi^{2}m_{e}^{16}}E^{10}+\ldots\label{A4}
\end{equation}

\begin{equation}
4\pi\frac{\partial\mathcal{L}^{(1)}}{\partial E}=\frac{2\alpha^{2}}{45\pi m_{e}^{4}}E^{3}+\frac{\alpha^{3}}{630\pi m_{e}^{8}}E^{5}+\frac{32\alpha^{4}}{315\pi m_{e}^{12}}E^{7}+\frac{160\alpha^{5}}{297\pi m_{e}^{16}}E^{9}.
\end{equation}

We list the coefficients in the Lagrangian expansion: 
\begin{eqnarray*}
a_0 & = & \frac{\alpha^{2}}{360\pi^{2}m^{4}}\\
a_1 & = & \frac{\alpha^{3}}{630\pi^{2}m^{8}}\\
a_2 & = & \frac{\alpha^{4}}{315\pi^{2}m^{12}}\\
a_3 & = & \frac{4\alpha^{5}}{297\pi^{2}m^{16}}\\
a_4 & = & \frac{22112\alpha^{6}}{225225\pi^{2}m^{20}}\\
a_5 & = & \frac{128\alpha^{7}}{117\pi^{2}m^{24}}\\
a_6 & = & \frac{462976\alpha^{8}}{26775\pi^{2}m^{28}}\\
a_7 & = & \frac{22459904\alpha^{9}}{61047\pi^{2}m^{32}}\\
a_8 & = & \frac{1430413312\alpha^{10}}{141075\pi^{2}m^{36}}\\
a_9 & = & \frac{20364132352\alpha^{11}}{8\times7245\pi^{2}m^{40}}\\
a_{10} & = & \frac{123922856542208\alpha^{12}}{8\times1036035\pi^{2}m^{44}}\\
a_{11} & = & \frac{1379781312512\alpha^{13}}{8\times225\pi^{2}m^{48}}\\
a_{12} & = & \frac{56921405366730752\alpha^{14}}{8\times152685\pi^{2}m^{52}}
\end{eqnarray*}
Note that the coefficients in Eq. (\ref{etacoeffs}) are approximately related to the 
coefficients above, i.e., $a_1 \approx \eta_1$ and $a_2 \approx \eta_2$.
From this we get the following coefficients in
the polynomial equation for the electric
field ($\sum_0\tilde{\eta}_i E^{2i+1}=E_0$ with $\tilde{\eta}_0=1$)
\begin{eqnarray*}
\tilde{\eta}_1 & = & 16\pi\times a_0\\
\tilde{\eta}_2 & = & 24\pi\times a_1\\
\tilde{\eta}_3  & = & 32\pi\times a_2\\
\tilde{\eta}_4  & = & 40\pi\times a_3\\
\tilde{\eta}_5  & = & 48\pi\times a_4\\
\tilde{\eta}_6  & = & 56\pi\times a_5\\
\tilde{\eta}_7  & = & 64\pi\times a_6\\
\tilde{\eta}_8  & = & 72\pi\times a_7\\
\tilde{\eta}_{9}  & = & 80\pi\times a_8\\
\tilde{\eta}_{10}  & = & 88\pi\times a_9\\
\tilde{\eta}_{11}  & = & 96\pi\times a_{10}\\
\tilde{\eta}_{12}  & = & 104\pi\times a_{11}\\
\tilde{\eta}_{13}  & = & 112\pi\times a_{12}
\end{eqnarray*}

From the Lagrangian (\ref{A4}), the symmetric energy-momentum tensor
can be calculated using the expression

\begin{equation}
T_{\mu\nu}=\frac{\partial\mathcal{L}}{\partial\mathcal{F}}F_{\mu}^{\:\alpha}F_{\nu\alpha}-g_{\mu\nu}\mathcal{L}.
\end{equation}

For the 00 component we have

\begin{equation}
T_{00} = \frac{b_{1}}{2}E^{2}+\frac{3}{4}b_{2}E^{4}+\frac{5}{8}b_{3}E^{6}+\frac{7}{16}b_{4}E^{8}+\frac{9}{32}b_{5}E^{10}+\ldots.\label{A8}
\end{equation}
with 
\begin{eqnarray*}
b_{1} & = & \frac{1}{4\pi}\\
b_{2} & = & \frac{\alpha}{90\pi^{2}m_{e}^{4}},\\
b_{3} & = & \frac{4\alpha^{3}}{315\pi^{2}m_{e}^{8}},\\
b_{4} & = & \frac{16\alpha^{4}}{315\pi^{2}m_{e}^{12}}.\\
b_{5} & = & \frac{128\alpha^{5}}{297\pi^{2}m_{e}^{16}}\\
b_{6} & = & \frac{1415168\alpha^{6}}{225225\pi m_{e}^{20}}
\end{eqnarray*}

\end{document}